\documentclass[a4paper,11pt]{article}
\usepackage{jcappub} 

\usepackage{bm}
\usepackage{epsfig}
\usepackage{longtable}
\usepackage{deluxetable}
\usepackage[normalem]{ulem} 
\usepackage{xcolor}

\newcommand{\RefMarkA}[1]{{#1}}
\newcommand{\RefDelA}[1]{{}}
\newcommand{\RefMarkB}[1]{{#1}}
\newcommand{\RefDelB}[1]{{}}
\newcommand{\RefMarkC}[1]{{#1}}
\newcommand{\RefDelC}[1]{{}}

\newcommand{\myNi}{\emph{(i)}\,}
\newcommand{\myNii}{\emph{(ii)}\,}
\newcommand{\myNiii}{\emph{(iii)}\,}

\newcommand{\mass}{\bar{m}}
\newcommand{\myeta}{{\eta}}
\newcommand{\mySagn}{s_p}
\newcommand{\myLAGN}{L_p}
\newcommand{\myEz}{\mathcal{H}}
\newcommand{\myTS}{\mbox{TS}}
\newcommand{\myR}{R_{500}}
\newcommand{\model}{\mu}
\newcommand{\Mbin}{\mathcal{M}}
\newcommand{\const}{\mbox{const}}
\newcommand{\gama}{$\gamma$}
\newcommand{\mytautheta}{\tau}
\newcommand{\mytaur}{\varrho}
\newcommand{\till}{{\mbox{--}}}

\newcommand{\ie}{\emph{i.e.} }
\newcommand{\eg}{\emph{e.g.,} }
\newcommand{\fin}{\mbox{ .}}
\newcommand{\coma}{\mbox{ ,}}

\newcommand{\cm}{\mbox{ cm}}
\newcommand{\sr}{\mbox{ sr}}
\newcommand{\se}{\mbox{ s}}
\newcommand{\yr}{\mbox{ yr}}

\newcommand{\erg}{\mbox{ erg}}

\newcommand{\km}{\mbox{ km}}

\newcommand{\kpc}{\mbox{ kpc}}
\newcommand{\Mpc}{\mbox{ Mpc}}
\newcommand{\eV}{\mbox{ eV}}
\newcommand{\keV}{\mbox{ keV}}
\newcommand{\MeV}{\mbox{ MeV}}
\newcommand{\GeV}{\mbox{ GeV}}
\newcommand{\TeV}{\mbox{ TeV}}

\newcommand{\K}{\mbox{ K}}
\newcommand{\dgr}{^{\circ}}

\newcommand{\Fukazawa}{\cite{FukazawaEtAl04}}
\newcommand{\Chen}{\cite{ChenEtAl07}}

\newcommand{\mynewcommand}[2]{\ifdefined #1 \else \newcommand{#1}{#2} \fi}
\mynewcommand\pasj{{PASJ}}   
\mynewcommand{\apj}{ApJ}     
\mynewcommand{\apjl}{ApJL}     
\mynewcommand{\apjs}{ApJS}    
\mynewcommand\mnras{{MNRAS}} 
\mynewcommand{\aap}{A\&A}    
\mynewcommand{\nat}{Nature}  
\newcommand\jcap{{JCAP}}  
\newcommand\nar{{New A ReV.}}  

\title{Detection of virial shocks in stacked Fermi-LAT galaxy clusters}

\author[a,b]{Ido Reiss}
\author[a]{Uri Keshet}

\affiliation[a]{Physics Department, Ben-Gurion University of the Negev, \\POB 653, Be'er-Sheva 84105, Israel}
\affiliation[b]{Physics Department, Nuclear Research Center Negev, \\POB 9001, Be'er-Sheva 84190, Israel}

\emailAdd{reissi@post.bgu.ac.il}
\emailAdd{ukeshet@bgu.ac.il}

\abstract{
	Galaxy clusters are thought to grow by accreting mass through large-scale, strong, yet elusive, virial shocks. Such a shock is expected to accelerate relativistic electrons, thus generating a spectrally-flat leptonic virial-ring.
	However, until now, only the nearby Coma cluster has shown evidence for a $\gamma$-ray virial ring.
	We stack \emph{Fermi}-LAT data for the 112 most massive, high latitude, extended clusters, enhancing the ring sensitivity by rescaling clusters to their virial radii and utilizing the expected flat energy spectrum.
    In addition to a central unresolved, hard signal (detected at the $\sim 5.8\sigma$ confidence level), probably dominated by AGN, we identify (at the $5.8\sigma$ confidence level) a bright, spectrally-flat $\gamma$-ray ring at the expected virial shock position.
    The ring signal implies that the shock deposits $\sim 0.6\%$ (with an interpretation uncertainty factor $\sim2$) of the thermal energy in relativistic electrons over a Hubble time.
    This result, consistent with the Coma signal, validates and calibrates the virial shock model, and indicates that the cumulative emission from such shocks significantly contributes to the diffuse extragalactic $\gamma$-ray and low-frequency radio backgrounds.
}
\begin{document}

\maketitle
\flushbottom

\section{Introduction}
\label{sec:Intro}

In the hierarchical paradigm of large-scale structure (LSS) formation, galaxy clusters are the largest objects ever to virialize. With a mass $M$ in excess of $10^{13}M_\odot$ or even $10^{15}M_\odot$, they are located at the nodes of the cosmic web, where they accrete matter from the surrounding voids and through large-scale filaments. Due to their vast size, galaxy clusters resemble island universes seen at great distance, providing a powerful cosmological probe and a unique astrophysical laboratory.

Galaxy clusters are thought to grow by accreting gas through strong, collisionless, virial shocks, surrounding each cluster.
These shocks form as the accreted gas abruptly slows down and heats to virial temperatures.
They mark the edge of the cluster, and could provide a wealth of information regarding structure formation, large-scale structure, and shock physics.
However, until now, no clear shock signal has been confirmed.

Strong collisionless shocks are thought, by analogy with supernova remnant (SNR) shocks, to accelerate charged particles to highly relativistic, $\gtrsim 10\TeV$ energies.
These particles, known as cosmic ray (CR) electrons (CREs) and ions (CRIs), are accelerated to a nearly flat, $E^2dN/dE\propto \const.$ spectrum (equal energy per logarithmic CR energy bin), radiating a distinctive non-thermal signature which stands out at the extreme ends of the electromagnetic spectrum, in particular as high energy \gama-rays.

High-energy CREs cool rapidly, on timescales much shorter than the Hubble time $H^{-1}$, by Compton-scattering cosmic microwave-background (CMB) photons
\cite{LoebWaxman00, TotaniKitayama00, KeshetEtAl03}.
These up-scattered photons should then produce \gama-ray emission in a thin shell around the galaxy cluster, as anticipated analytically \cite{WaxmanLoeb00, TotaniKitayama00} and calibrated using cosmological simulations \cite{KeshetEtAl03, Miniati02}.
The projected \gama-ray signal typically shows an elliptic morphology, elongated towards the large-scale filaments feeding the cluster \cite{KeshetEtAl03,KeshetEtAl04}.

The estimated \gama-ray luminosity of the virial shock scales as $L_s\propto \dot{M}T\propto M^{5/3}$, where $\dot{M}$ and $T\propto M^{2/3}$ are the mass accretion rate and temperature of the cluster.
The signal is therefore thought to be strongest in massive, hot, strongly accreting clusters.
The same \gama-ray emitting CREs are also expected to generate an inverse-Compton ring in hard X-rays \citep{KushnirWaxman10}, soft X-rays \cite{KeshetReiss17} and optical bands \citep{YamazakiLoeb15}, and a synchrotron ring in radio frequencies \citep{WaxmanLoeb00, KeshetEtAl04, KeshetEtAl04_SKA}.
These rings should coincide with a cutoff on the thermal Sunyaev-Zel'dovich (SZ) signal, marking the pressure drop beyond the shock, \ie at larger radii \cite{KocsisEtAl05}.

Once the energy accretion rate $\sim k_BT\dot{M}/\mass\propto \dot{M}T$ of the cluster has been determined, for example using an X-ray-calibrated isothermal $\beta$-model, its \gama-ray signature depends on a single free parameter, namely the CRE acceleration efficiency $\xi_e$, defined as the fraction of downstream energy deposited in CREs.
Here, $\mass$ is the mean particle mass and $k_B$ is the Boltzmann constant.
As high-energy CREs are short lived, the \gama-ray signal should reflect their spatially- and temporally-variable injection rate.
Locally, the signal thus depends on the single free parameter $\xi_e\dot{m}$, where $\dot{m}\equiv \dot{M}/(MH)$ is the dimensionless mass accretion rate and $H$ is Hubble's constant.

A promising target in the search for such a signal is the Coma cluster, as it is nearby, hot, massive, and found in the low-foreground region near the north Galactic pole.
An analysis \citep{KeshetEtAl12_Coma} of a $\sim220\GeV$ VERITAS mosaic of Coma \citep{VERITAS12_Coma} found evidence for a large-scale, extended \gama-ray feature surrounding the cluster.
The signal is best described as an elongated, thick, elliptical ring, with semi-minor axis coincident with the cluster's virial radius, oriented towards the LSS filament connecting Coma with Abell 1367.
The signal is seen at a nominal $2.7\sigma$ confidence level, but there is substantial evidence that it is real.
This includes a higher, $5.1\sigma$ significance found when correcting for the observational and background-removal modes, indications that an extended signal was indeed removed by the background model, correlations with synchrotron and SZ tracers, good agreement ($3.7\sigma$) with the simulated ring of the cluster, and the absence of extended signal tracers in other VERITAS mosaics.

Subsequent attempts to measure the \gama-ray signal from Coma using the \emph{Fermi} Large Area Telescope (LAT; henceforth) have failed, largely because it is difficult to reach the combined high sensitivity, controlled foreground, good resolution, and high --- yet not too high --- energy, set by VERITAS.
For example, broad band, $>100\MeV$ analyses \citep{ZandanelAndo14, Prokhorov14} of LAT data found no excess emission from Coma, placing upper limits $\xi_i<15\%$ on CRI acceleration and $\xi_e<1\%$, and questioning spectrally-flat emission matching the VERITAS signal. However, at these low energies, the point spread function (PSF) is prohibitively large \citep{AtwoodEtAl13}, with $68\%$ ($95\%$) containment exceeding $5\dgr$, far beyond (exceeding $13\dgr$, an order of magnitude above) the $1.3\dgr$ virial radius.
In addition, such upper limits are sensitive to the morphology of the modeled signal, which is not well-constrained by the observational mode that generated the VERITAS mosaic.
Moreover, an extended LAT signal around Coma was later reported \citep{FermiComa16}, partly overlapping the virial radius.
This signal, still below the threshold needed to claim LAT detection, is consistent with the VERITAS signal when correcting for the larger extent of the latter.

A follow-up search \cite{KeshetReiss17} in $1\till30\GeV$ \gama-rays from the LAT using a thin elliptical template did find a $3.4\sigma$ LAT excess, at the ring elongation and orientation inferred from VERITAS.
A corresponding template applied to soft, $\sim0.1\keV$ X-rays from \emph{ROSAT} bands R1+R2 discovered ($>5\sigma$) the expected signature from the same ring parameters \cite{KeshetReiss17}.
The significances of both LAT and \emph{ROSAT} signals are maximal near the VERITAS ring parameters. The intensities of the \emph{ROSAT}, LAT, and VERITAS signals are consistent with a virial shock with a CRE acceleration rate $\xi_e\dot{m}\sim0.3\%$ (with an uncertainty factor of $\sim3$) and a nearly flat, $p\equiv−d\ln N_e/d\ln E \simeq 2.0\till2.2$ spectrum.
The evidence, combined, indicates a high confidence detection of an elongated virial ring.
The sharp radial profiles of the LAT and \emph{ROSAT} signals suggest preferential accretion in the plane of the sky, as indicated by the distribution of neighboring large-scale structure.

An alternative approach to focusing on a single cluster is to boost the virial shock signal by stacking the data of many different clusters.
By correlating the EGRET data with 447 rich ($R\geq2)$ Abell clusters, a possible association of \gama-ray emission with clusters was reported at the $3\sigma$ confidence level \cite{ScharfMukherjee02}.
In comparison with source number counts computed analytically \cite{WaxmanLoeb00} and numerically \cite{KeshetEtAl03}, it correspond to an average $\xi_{e}\dot{m}\simeq 4\%$.
However, the morphology of the signal is unclear due to the low resolution of EGRET, and its association with virial shocks is unlikely given the strong signal and the low typical mass ($\sim10^{13}M_\odot$) of the clusters in the sample.
A subsequent search \cite{ReimerEtAl03} for a correlation between EGRET data and 58 X-ray bright clusters showed no signal.
Later attempts to stack the higher sensitivity, better resolved LAT data \cite{ReimerEtAl03,AckermannEtAl10, AckermannEtAl14_GammaRayLimits, HuberEtAl13, ProkhorovChurazov14, ZandanelAndo14, GriffinEtAl14} failed to identify a diffuse signal, thus questioning the validity of the results of Ref. [\citenum{ScharfMukherjee02}].

LAT data for a sample of 50 clusters did suggest ($2.7\sigma$) excess emission from the core regions, but this was identified as emission from unknown (at the time) point sources in three individual clusters \cite{AckermannEtAl14_GammaRayLimits}.
Stacking 55 X-ray bright clusters on the same angular scale, \ie without rescaling their images, indicated ($4.3\sigma$) excess \gama-ray emission from the central $r=0.25\dgr$ radius of the clusters \cite{ProkhorovChurazov14}, but this was attributed to emission from active galactic nuclei (AGN).
Stacking 78 rich clusters on the same distance scale placed an upper limit ($95\%$ confidence level) of $2.3\times10^{-11} \se^{-1}\cm^{-2}$ on the $(0.8-100)\GeV$ flux \cite{GriffinEtAl14}, but the $2\Mpc$ resolution used is comparable to the typical virial radius of the clusters, and the limit pertains to $r<2\Mpc$, insufficient for very massive clusters or for an elliptical shock.
Correlations at $\sim2.7\till5.0\sigma$ confidence levels were reported between LAT data and three different galaxy cluster catalogs \cite{BranchiniEtAl17}, corroborated by stacking the \gama-rays on a fixed angular scale; however, these signals pertain to relatively high, $z\sim0.2$--$0.4$ redshifts, extend to large, few $10\Mpc$ scales far beyond the virial radius, and probably reflect a population of AGN and star-forming galaxies.

We stack the LAT \gama-ray emission from galaxy clusters, specifically targeting the virial rings.
Cluster virial radii span a wide range of angular and spatial scales; a ring signal would be smeared out by stacking data on either scale.
Hence, unlike previous studies, we stack the data of each cluster normalized to its virial radius.
We also utilize the nearly flat spectrum, by co-adding the independent photon counts in different energy bands.
The resulting high sensitivity is sufficient for picking up a ring signal at the expected position of the virial shock.
In addition, we find a hard unresolved signal from the center of the clusters, likely to arise from faint AGN.

The paper is organized as follows.
In \S\ref{sec:Data} we describe the preparation of the LAT data and the cluster sample used in this work.
The data stacking procedure is detailed in \S\ref{sec:simple}, where it is shown to robustly reveal both a peripheral signal and a central signal.
These signals are modeled as leptonic emission from the virial shock and a central point source in \S\ref{sec:TS}.
The results are summarized and discussed in \S\ref{sec:Discussion}.
We introduce our $\beta$ model-based analysis of virial-shock and AGN emission in Appendix \S\ref{sec:BetaModel}.
The parameters of the clusters in our sample are provided in \S\ref{sec:ClusterSample}.

We adopt a flat $\Lambda$CDM cosmological model with a Hubble constant $H_0 = 70 \km \se^{-1}$ $\Mpc^{-1}$ and a mass fraction $\Omega_m=0.3$.
Assuming a $76\%$ hydrogen mass fraction gives a mean particle mass $\mass\simeq 0.59m_p$.
Confidence intervals quoted are $68\%$ for one parameter; multi-parameter intervals are specified when used.

\section{Data preparation}
\label{sec:Data}
\begin{figure*}[ht]
	\centerline{\epsfxsize=16.8cm \epsfbox{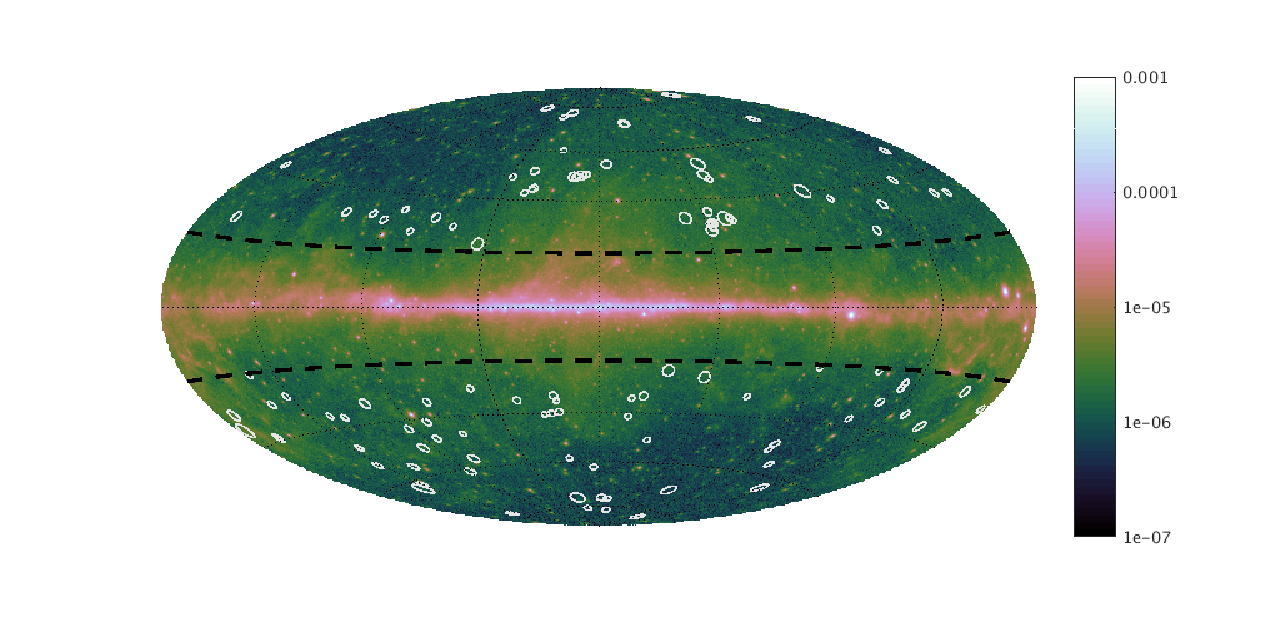}
		\vspace{-1cm}}
	\caption{\label{fig:cluster mapA}
		{\bf \emph{Fermi}-LAT sky map.}
		\emph{Fermi}-LAT photon flux (colorbar in units of $\se^{-1} \cm^{-2} \sr^{-1}$) sky map in the $(1\till500)\GeV$ energy range, shown in a Hammer-Aitoff projection with Galactic coordinates. The locations and enlarged sizes (white circles with $5\myR$ radii) of the 112 clusters used in the analysis are superimposed. Latitudes are shown every $20\dgr$, and longitudes every $45\dgr$ (dotted lines).	
	}
\end{figure*}

We use the archival, $\sim8$ year, Pass-8 LAT data from the Fermi Science Support Center (FSSC)\footnote[1]{\texttt{http://fermi.gsfc.nasa.gov/ssc}}, and the Fermi Science Tools (version \texttt{v10r0p5}).
Pre-generated weekly all-sky files are used, spanning weeks $9\till422$ for a total of $414$ weeks ($7.9\yr$), with ULTRACLEANVETO class photon events.
A zenith angle cut of $90\dgr$ was applied to avoid CR-generated $\gamma$-rays originating from the Earth's atmospheric limb, according to the appropriate
FSSC Data Preparation recommendations.
Good time intervals were identified using the recommended selection expression \texttt{(DATA\_QUAL==1) and (LAT\_CONFIG==1)}.
The resulting sky map is shown in Figure \ref{fig:cluster mapA}.

Sky maps were discretized using a HEALPix scheme \citep{GorskiEtAl05} of order $N_{hp}=10$, providing a mean $\sim 0.057\dgr$ pixel separation.
This is sufficient for analyzing virial rings of $\gtrsim 0.2\dgr$ scales with each pixel approximated as a point.
This $0.2\dgr$ scale, in turn, is chosen according to the high-energy PSF of the LAT, as it corresponds to the $68\%$ containment angle at a photon energy $E\gtrsim10$ GeV \citep{AtwoodEtAl13}.
Our results change modestly when lowering the HEALPix order to $N_{hp}=9$, and are converged for $N_{hp}>9$.

Event energies were logarithmically binned into $N_\epsilon=4$ energy bands in the (1--100) GeV range.
Point source contamination was minimized by masking pixels within the $90\%$ containment area \RefMarkA{(for each energy band)} of each point source in the LAT 4-year point source catalog (3FGL) \cite{FermiPSC}.
In order to reduce the Galactic foreground, we mask $|b|<20\dgr$ latitudes, near the \gama-ray-bright Galactic plane.

We stack the LAT data around a sample of clusters selected from the Meta-Catalog of X-ray Clusters (MCXC) \cite{PiffarettiEtAl11}.
In addition to the location of each cluster on the sky and the cluster mass $M_{500}$, the catalog specifies the redshift $z$ and radius $R_{500}$ of each cluster, so the corresponding angular radius $\theta_{500}$ can be computed.
Here, $\delta=500\delta_{500}$ is the over-density parameter, defining a radius $r_\delta$ and an enclosed mass $M_\delta$, such that the mean enclosed mass density $M_\delta/[(4\pi/3)r_\delta^3]$ is higher by a factor of $\delta$ than the critical mass density $\rho_c(z)$ of the Universe at redshift $z$; the value of $\delta$ is used as a subscript for $r$, $M$, and $\delta$ itself.

In order to construct an optimal sample of clusters for a \gama-ray ring search, we apply the following cuts to the catalog, selecting only clusters that satisfy all of the following criteria.
\begin{enumerate}
	\item
	Massive clusters: a mass $M_{500}>10^{13}M_\odot$ enclosed within $\myR$.
	\item
	Resolvable ring: an angular radius $\theta_{500}>0.2\dgr$, chosen according to the high-energy LAT PSF.
	\item
	Avoiding the Galactic plane: clusters located sufficiently far from the Galactic plane, with latitude $|b|>20\dgr$.
	\item
	Avoiding contamination by point sources: a distance of at least $1.8\dgr$ (the $90\%$ containment angle at $1\GeV$) from any 3FGL point source.
	\item
	Avoiding excessively extended, bright clusters: we avoid the 4 clusters with $\theta_{500}>0.5\dgr$ \RefMarkA{(Coma, A3526, NGC5813, NGC4636)}, which are too bright and too extended for our analysis. In particular, the \gama-ray foreground estimation (see \S\ref{sec:simple}) around such clusters would be sensitive to the method used.
\end{enumerate}

Out of the 1743 clusters in the MCXC catalog, the above cuts leave only 112 clusters, listed in Table \ref{tab:clusters}.
The locations and (over-sized, for illustrative purposes) spatial extents of these clusters are shown as circles in Figure \ref{fig:cluster mapA}, superimposed on the (1--500) GeV LAT sky map.
Note that the highly-extended Coma cluster is removed from our sample due to the last cut, so our analysis is independent of the \gama-ray signal already suggested by VERITAS and LAT data \cite{KeshetReiss17}.

\section{Stacking analysis}
\label{sec:simple}

Cluster virial shocks are expected to form at radii $R_s$ near or beyond $R_{200}\simeq 1.6 R_{500}$.
Numerical simulations suggest (\eg Eq. 4 in Ref. \citenum{KeshetEtAl04}) a mean scaled shock radius $\mytaur_s\equiv R_s/R_{500} \simeq 2.8$, reflecting elliptical shocks spanning a typical range $(1.9\till 3.8)R_{500}$.
In terms of (proper) spatial scales, virial shocks span a wide range of radii, due to the diversity in cluster parameters.
The radial dispersion is even more severe in terms of angular scales, due the wide range of cluster distances.
Therefore, unlike previous studies, we stack the data on the same scaled radius, defined as $\mytautheta\equiv \theta/\theta_{500}$.

The foreground, after point sources and the Galactic plane were masked, varies mainly on scales much larger than the anticipated extent of the cluster signal. Therefore, this remaining foreground can be accurately approximated using a polynomial fit on large scales. For each cluster, we thus consider an extended, $0<\mytautheta<\mytautheta_{max}\equiv 15$ disk region around its center, and fit the corresponding LAT data by an order $N_f=4$ polynomial in the angular coordinates $\mytautheta_x$ and $\mytautheta_y$.
This is done separately for each of the four energy bands.
Then, for each cluster $c$, each photon energy band $\epsilon$, and each radial bin centered on $\mytautheta$ with width $\Delta \mytautheta=0.5$, we define the excess emission $\Delta n\equiv n-f$ as the difference between the number $n$ of detected photons and the number $f$ of photons estimated from the fitted foreground.

\begin{figure}[t]
	\centerline{\epsfxsize=14cm \epsfbox{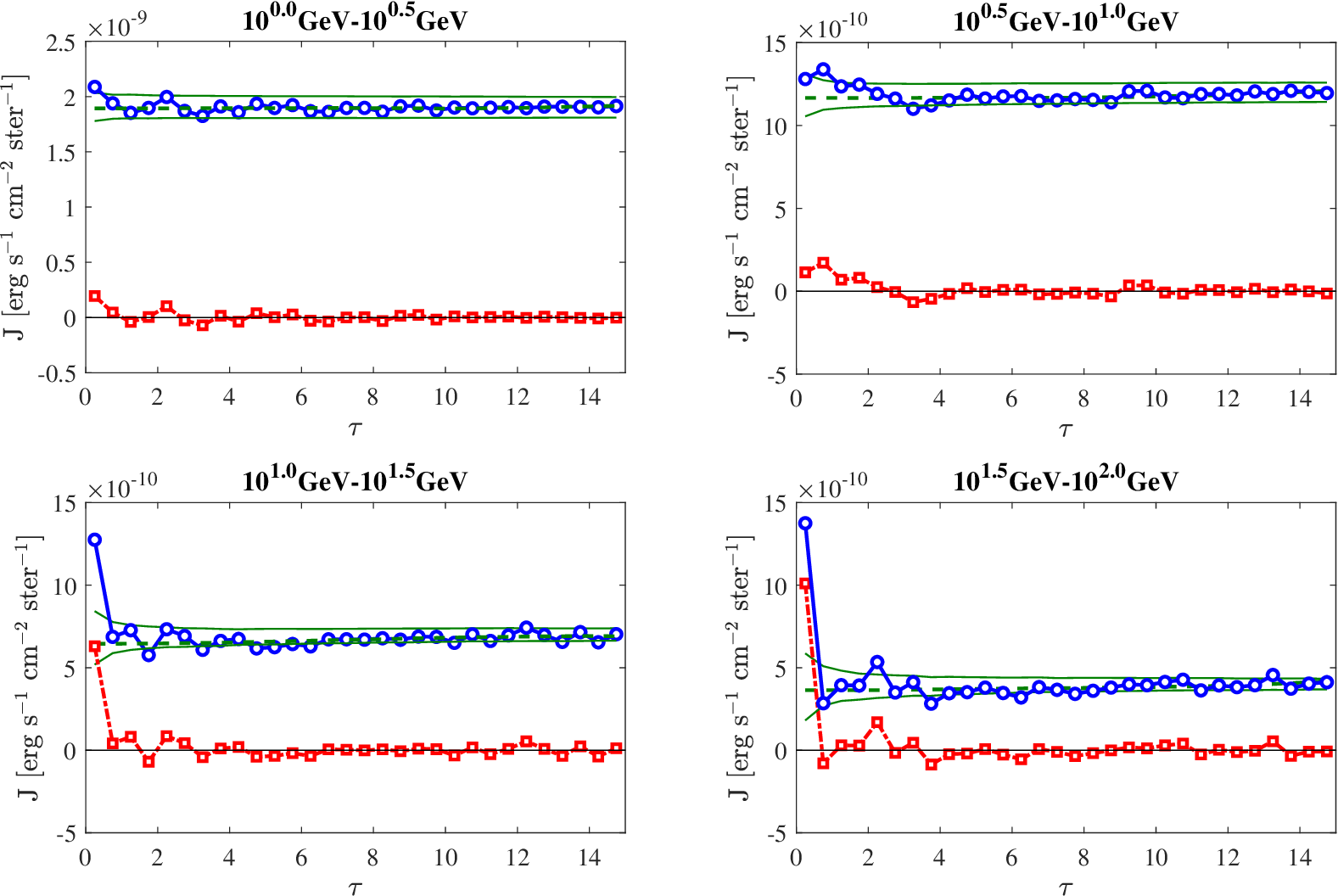}}
	\caption{\label{fig:flux} \RefMarkA{{\bf Energy flux of the stacked signal.} Scaled and radially binned energy flux, stacked over the cluster sample, shown as a function of the scaled angular radius $\mytautheta\equiv\theta/\theta_{500}$, in each of the four energy bands (blue circles, with solid lines to guide the eye). The estimated foreground is based on a fourth-order polynomial fit (henceforth; green dashed curves). The excess emission (red squares, with dash-dotted lines to guide the eye) suggests two signals: one in the central, $\tau<0.5$ bin (central signal) and one in the $2<\tau<2.5$ bin (peripheral signal).
}}
\end{figure}

The energy flux of the rescaled and radially stacked data is shown, for each of the four logarithmically spaced energy bands, as symbols in Figure \ref{fig:flux}. Also shown are the stacked fluxes that correspond to the estimated foreground (as dashed lines) and to the excess emission (as lower symbols).

The significance of the excess emission in a given cluster $c$, energy band $\epsilon$, and radial bin $\mytautheta$ can be estimated, assuming Poisson statistics with $f\gg1$, as
\begin{equation} \label{eq:SingleBinSignificance}
\nu_{\sigma,c}(\epsilon,\mytautheta) \simeq \frac{\Delta n_c}{\sqrt{f_c}} \fin
\end{equation}
Note that this estimate is undefined in regions where the foreground fit $f$ erroneously becomes non-positive.
However, these region are very rare; they appear only in the highest energy, photon-sparse band, and even there they constitute only $0.3\%$ of the radial bins.

Next, we stack the data over the clusters in the sample.
To examine the robustness of our analysis and possible biases by a large number of photons arriving from a few high-foreground or bright clusters, or from a high significance signal arriving from a few low foreground clusters, we define two different methods to compute the significance of the signal stacked over clusters.

The first, more standard method is photon co-addition.
Here, at a given radial bin and energy band, we separately sum the excess photon count and the foreground photon count over the $N_c$ clusters.
The stacked significance is evaluated as the ratio between the stacked excess and the square root of the stacked foreground,
\begin{equation}
\nu_\sigma^{(ph)}(\epsilon,\mytautheta) = \frac
{\sum_{c=1}^{N_c} \Delta n_c}
{\sqrt{\sum_{c=1}^{N_c} f_c}} \fin
\end{equation}
The second method is cluster co-addition.
Here, at a given radial bin and energy band, we co-add the significance $\nu_{\sigma,c}$ of Eq.~(\ref{eq:SingleBinSignificance}) over the $N_c^*(\epsilon,\mytautheta)$ clusters for which it is defined (\ie where $f_c>0$),
\begin{equation} \label{eq:clusterCoAddition}
\nu_\sigma^{(cl)}(\epsilon,\mytautheta) = \frac
{\sum_{c=1}^{N_c^*} \nu_{\sigma,c}}
{\sqrt{N_c^*}} \fin
\end{equation}
The two methods qualitatively agree with each other, although they do differ in a handful of bins by up to $\sim 1\sigma$. The difference between the two methods gauges the stacking systematics.

In both methods, we next co-add the $N_\epsilon=4$ logarithmic energy bands with equal weights,
\begin{equation} \label{eq:BandCoAdd}
\nu_\sigma(\bm{\epsilon},\mytautheta) = \frac
{\sum_{\epsilon=1}^{N_\epsilon}\nu_{\sigma}(\epsilon,\mytautheta)}
{\sqrt{N_\epsilon}} \fin
\end{equation}
The vector $\bm{\epsilon}$ indicates the co-addition of energy bands, while index $\epsilon$ pertains as above to a specific energy band.

\begin{figure}[t]
	\centerline{
		\epsfxsize=14cm \epsfbox{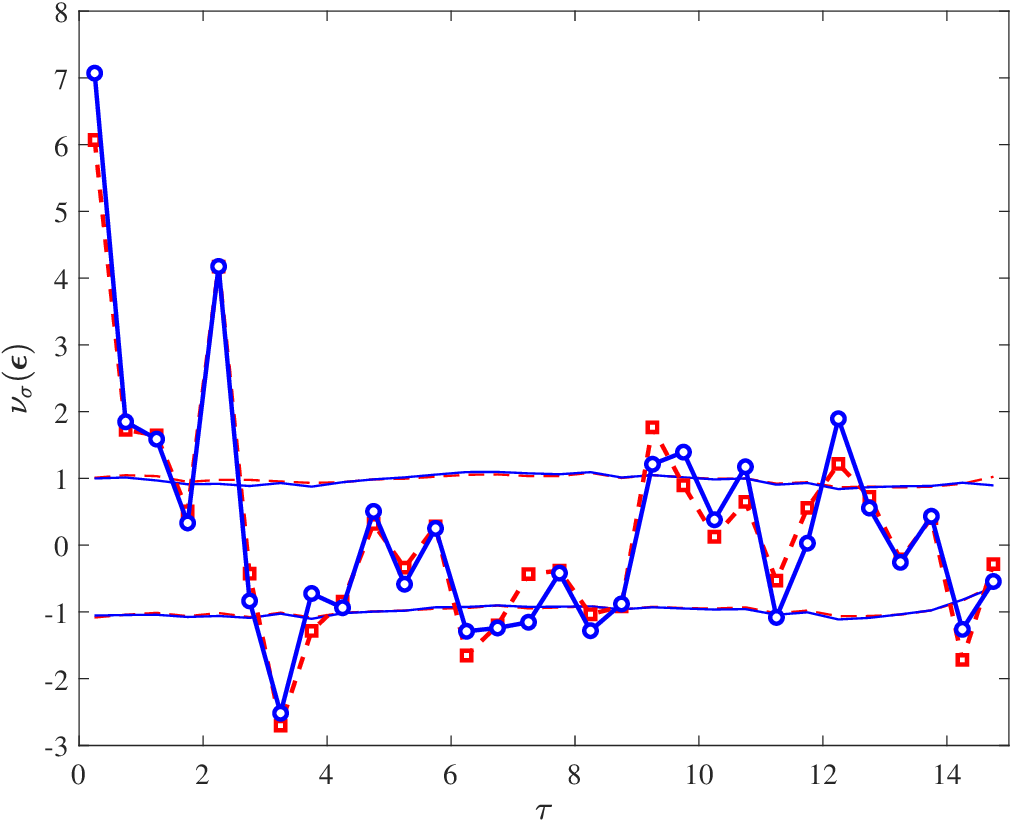}
	}
	\caption{\label{fig:combined1}
		\RefMarkA{{\bf Significance of the stacked signal excess.}
		The significance $\nu_\sigma(\bm{\epsilon})$ of the energy co-added excess \gama-ray counts over the foreground, shown as a function of $\mytautheta$. The excess was stacked over the cluster sample both by photon co-addition (blue circles with solid lines to guide the eye) and by cluster (\ie per-cluster significance) co-addition (red rectangles with dashed lines).
		The $1\sigma$ extents of the mock catalog distributions are shown (thin lines) for photon co-addition (solid blue curve) and for cluster co-addition (dashed red).
	}}
\end{figure}

The resulting significance of the excess emission is shown in Figure \ref{fig:combined1}.
Both the flux (Figure \ref{fig:flux}) and the significance (Figure \ref{fig:combined1}) show two spatially separated components: a central component and a peripheral, ring-like component.
The central emission is unresolved, confined to the innermost $\mytautheta<0.5$, where it presents at an energy co-added significance of $6\sigma\till7\sigma$.
It is morphologically consistent with a point source located at the center of the cluster.
The peripheral, ring-like signal peaks at $2.0<\mytautheta<2.5$, where it presents at a significance of $4.2\sigma$.
This signal matches the expected signature of \gama-ray rings arising from inverse-Compton scattering of CMB photons by virial-shock accelerated CREs.

\begin{figure*}[t]
	\centerline{
    	\epsfxsize=8cm \epsfbox{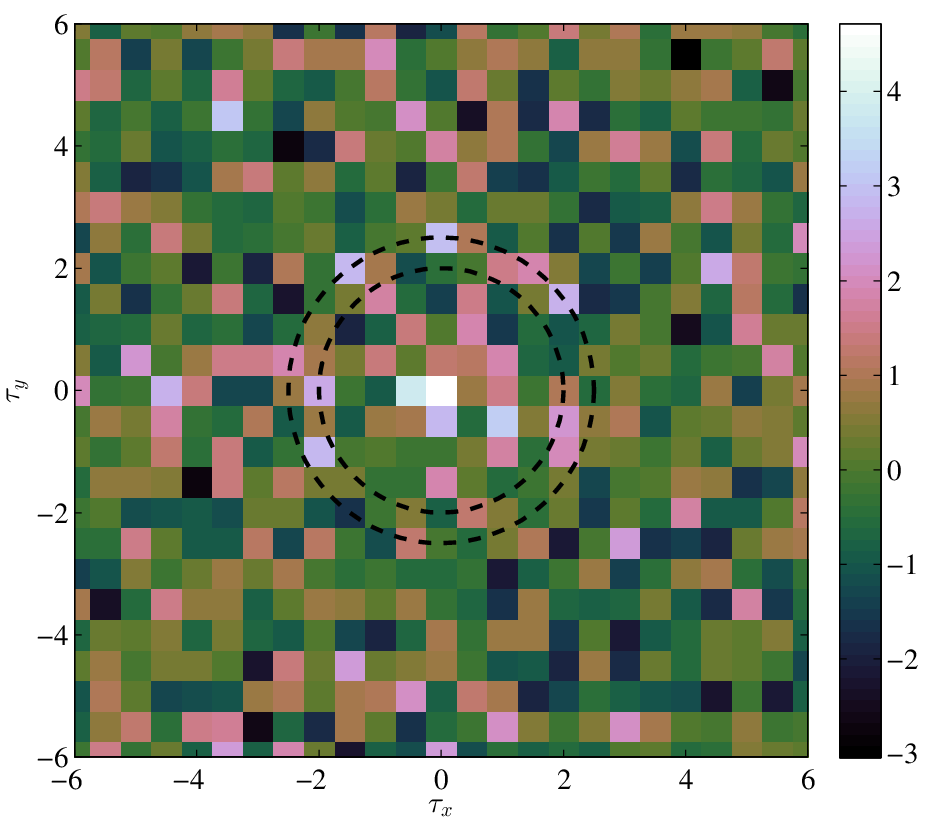}
		\epsfxsize=8cm \epsfbox{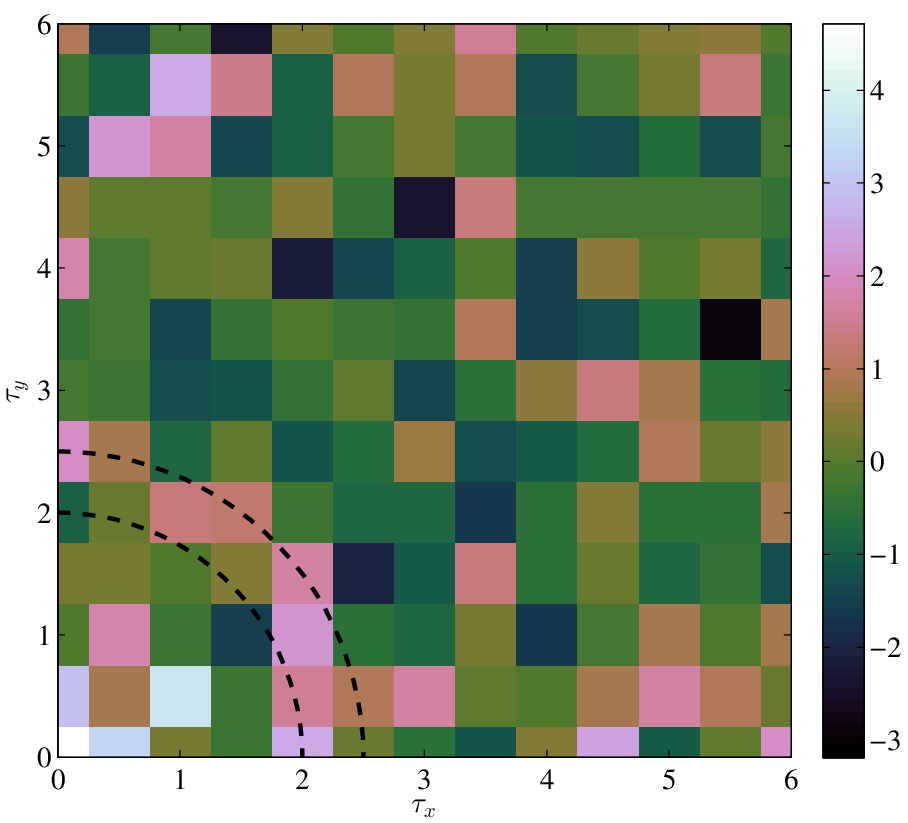}
}
	\caption{\label{fig:sig_mapA}
		{\bf Stacked images of the excess significance.}
		The significance of the stacked \gama-ray excess (colorbar: $\nu_\sigma(\bm{\epsilon})$ with a cubehelix color map, Ref. \citenum{Green11_Cubehelix}), after photon and energy co-addition. The image is shown in the normalized $\mytautheta_x$--$\mytautheta_y$ plane centered on the clusters (left), and four-folded onto one quadrant (right).
        The dashed quadrant circles (left) or arcs (right), shown as a guide to the eye, enclose the $2.0<\mytautheta<2.5$ radial bin of the peripheral signal.
		Folding onto a quadrant is necessary in order to render the ring marginally visible by eye. Note that the highest significance pixel lies at the very center ($\mytautheta<0.5$).
	}
\end{figure*}

The two components, each arising from the cumulative contribution of many clusters, are found in a wide range of cluster masses.
The signals are marginally discernable in the four-folded, stacked image, shown in Figure \ref{fig:sig_mapA} with a guide to the eye, even without radial binning; the central signal is seen even without such folding.

In order to validate the foreground-based significance estimation and to examine possible systematic biases, we prepare and analyze a large number \RefDelB{($N_{mock}$=$2000$) }of control, \ie mock, cluster catalogs.
\RefMarkB{We use $N_{mock}=20,\!000$ catalogs, approximately saturating the number of independent samples.}
In each mock catalog we use the exact same cluster masses and angular radii $\theta_{500}$ as those in the true sample, but place the mock clusters in random yet constrained locations on the sky. The constraints assure that the mock clusters satisfy the same cut criteria of the true cluster sample, avoiding the Galactic plane and point source contamination in the same way as the real clusters do.

For a large enough mock sample, well-behaved data, and a good foreground determination, one expects the significance of the mock excess counts to converge on a mean $\langle \nu_\sigma\rangle\to 0$ and a variance $\mbox{Var}(\nu_\sigma)\to 1$.
Such a behavior, supplemented when possible by higher moments of the mock catalog that follow a normal distribution, would support the interpretation of the estimated significance as correctly representing a normal distribution.

The $1\sigma$ band of the mock clusters is shown in Figure \ref{fig:combined1} as thin lines (a pair of thin lines for each co-addition method). The mean $\langle \nu_\sigma\rangle$ of the mock catalogs deviates from zero by no more than $0.1$, revealing no large systematic bias. The variance $\mbox{Var}(\nu_\sigma)$ of the mock catalogs deviates appreciably from unity only beyond $\mytautheta\simeq 12$, suggesting that out to this large radius (where the foreground estimate is expected to become less accurate), our significance estimates are reliable. \RefMarkB{Using the standard deviation inferred from the mock sample, rather than from Poisson statistics, would change the nominal significance by less than $2\%$.}

\begin{figure}[t]
	\centerline{\epsfxsize=8.4cm \epsfbox{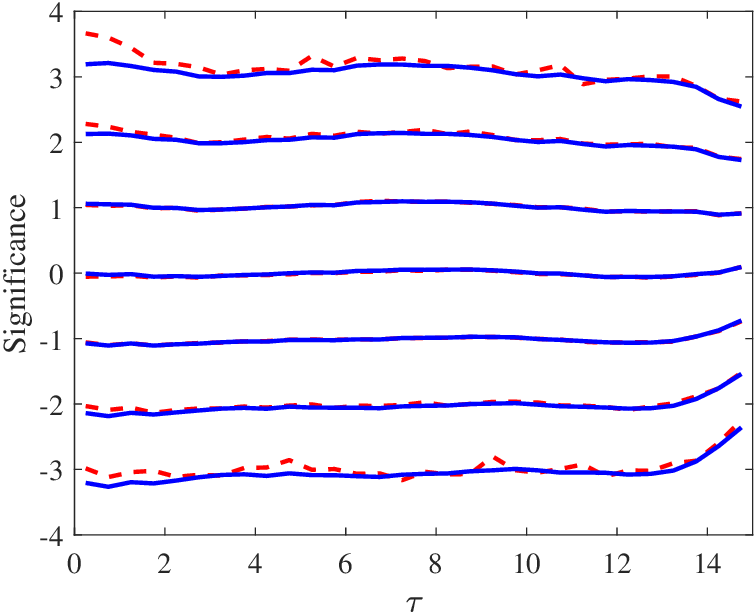}}
	\caption{\label{fig:check_normal} \RefMarkB{{\bf Mock catalogs distribution.}} The symmetric $68\%$, $95\%$, $99.7\%$ and confidence intervals of the photon and energy co-added significance of the excess counts, as a function of $\mytautheta$, inferred from the mock catalogs (dashed red).
		These are compared with the standard deviation of the mock sample, multiplied by $\pm1$, $\pm2$, and $\pm3$ and added to the mock mean (solid blue).
		The agreement suggests that a normal distribution can be assumed at least out to $\pm3\sigma$.
	}
\end{figure}

The validity of the foreground-based significance estimation is further supported by confirming that the mock significance approximately follows the expected normal distribution out to at least the $\pm3\sigma$ confidence level\RefMarkB{. In Figure \ref{fig:check_normal}, the $68\%$, $95\%$, $99.7\%$ confidence intervals from the mocks are compared to the $\pm1,2,3$ standard deviation intervals.
The agreement between these curves indicates that the mock distribution is consistent with an unbiased normal distribution.}

\RefDelB{
The angular separation between the clusters in our sample is on average sufficiently large to avoid an overlap between their individual regions of interest.
However, two regions on the sky show a high density of clusters, where the regions of interest overlap and some of the photons are double-counted.
One region contains 12 clusters around coordinates $\{l,b\}\simeq \{315\dgr,32\dgr\}$, and the other contains four clusters around coordinates $\{12\dgr,50\dgr\}$. Removing these 16 clusters from our sample strengthens (see Figure \ref{fig:dip_removed}) the $2<\tau<2.5$ signal to $\sim4.4\sigma$, while diminishing the $\tau<0.5$ signal by $\sim0.2\sigma$.}

\RefDelA{
The sensitivity of the results to sub-threshold, undetected point sources tested by using the 2-year LAT point source catalog (2FGL) \cite{Fermi2FGL} for masking, instead of 3FGL. This earlier catalog was created using data with approximately half the exposure of 3FGL. Only slight changes in the significance of both the peripheral signal (no effect for the cluster co-addition, and a $\sim0.2\sigma$ reduction for the photon co-addition) and the central signal (a $\sim0.1\sigma$ reduction).
These results confirm that our conclusions are not sensitive to unresolved point sources.
}

\RefDelB{
By examining cluster sub-samples with increasingly narrower ranges of $\theta_{500}$, we test for systematic effects that may arise from the radial $\mytautheta$ rescaling, and for the statistical behavior of the stacked signals. We find that the significance of the signals indeed scale as expected from the statistics of the number of clusters in the sub-sample. For example, for the peripheral signal, where the full 112 cluster sample gives $\nu_\sigma(\bm{\epsilon})\simeq4.2$, the 83 clusters with $0.2\dgr<\theta_{500}<0.3\dgr$ give $\nu_\sigma(\bm{\epsilon})\simeq3.7$, and the 62 clusters with $0.2\dgr<\theta_{500}<0.25\dgr$ give $\nu_\sigma(\bm{\epsilon})\simeq3.0$, as expected from Poisson statistics.
This confirms that the rescaling of the cluster angular sizes does not introduce significant spurious systematic effects, and that the stacked signals arise from the cumulative contribution of many clusters, and are statistically limited.
}

We carry out a suite of convergence and sensitivity tests, indicating that our results are robust to variations in the preparation of the LAT data (masking of point sources and the Galactic plane) and of the cluster sample (mass cuts, $\theta_{500}$ cuts, \RefMarkB{overlap between cluster regions of interest,} and proximity to point sources and to the Galactic plane), in the photon analysis methods (discretization, foreground modeling), in the cluster stacking methods (rescaling to $\theta_{500}$, photon vs. \RefDelB{per }cluster significance co-addition, different mass bin co-additions, radial bin size), and in our energy co-addition method (number of energy bins). \RefMarkB{We demonstrate these tests in Appendix \ref{sec:sensitivity}.}

As in other cases where the foreground determination is influenced by the signal itself, the above results underestimate the true significance of the signal, by $\sim40\%$ according to control catalogs. Nevertheless, as shown below, the signal parameters are robust to the foreground subtraction.

\section{Modeling}
\label{sec:TS}

In order to analyze the excess emission, and to accurately determine the significance of the signals, we present a model for each component.
For the ring-like emission, a simple model, based on a spherical, isothermal, $\beta$-model gas distribution, is presented in Appendix \ref{sec:BetaModel}. Once the gas distribution in a cluster has been determined --- here using the tabulated $\beta$-model parameters based on the X-ray signature of the cluster --- the leptonic model essentially has only two free parameters: the CRE acceleration efficiency, $\xi_{e}$ and the virial shock scaled radius $\mytaur_{s}$ (see Appendix \ref{sec:RingModel}). For the central component, we examine a simple point source model (see Appendix \ref{sec:AGNmodel}) with luminosity $\myLAGN$ and spectral index $\mySagn$. \RefMarkB{These four parameters are thought to depend only weakly on the redshift. As our sample is restricted to the narrow range $z<0.1$, we approximate these parameters as redshift independent.}

Out of the 112 clusters in our MCXC-based sample, 44 clusters have been fit with a $\beta$-model \cite{FukazawaEtAl04, ChenEtAl07}, such that the density profile index $\beta$ and the temperature $T$ of the gas are approximately determined. In clusters with an unknown $T$, we use the mass-temperature relation (Eq. (\ref{eq:M_T relation})) implied by hydrostatic equilibrium to compute the flux. In clusters with an unknown $\beta$, we adopt the mean value inferred from the other clusters in each mass bin (see below).

We evaluate the model parameters and their uncertainty, taking into account the PSF corrections, the signal and foreground photon statistics, and the correlations that are induced by cuts in the map, by masked pixels, and by our methods of stacking.
This is done using control samples, each of which Monte Carlo simulates the LAT data that would arise from the clusters of a mock catalog for a given choice of model parameters.
The resulting mock photon counts are then injected into the real LAT data, and the result is analyzed with the same pipeline used to study the real clusters.
We repeat this for $N_{mock}=10$ catalogs, and for a large set of parameter values.
Each mock cluster corresponds to a real cluster in our sample, and is assigned with the same parameters but with a random location in the permitted region of the sky.

A maximal likelihood (minimal $\chi^2$) analysis is used to calibrate the model and estimate the uncertainties in the parameters.
First, for given $\epsilon$ bin, $\mytautheta$ bin, and mass bin \RefMarkB{$\Mbin=\{1,\ldots,n_M\}$,}
we compute the $\chi^2$ of the excess counts $\Delta n_c(\epsilon,\mytautheta,\Mbin)$ in the real sub-sample, with respect to the model prediction $\model_c(\epsilon,\mytautheta,\Mbin)$, both quantities photon co-added over the clusters $c$ in the mass bin,
\begin{equation} \label{eq:ChiSquared}
\chi^2(\epsilon,\mytautheta,\Mbin)=\frac{\left[\sum_{c}\left(\Delta n_c - \model_c\right)\right]^2}{\sum_c\left(f_c + \model_c\right)} \fin
\end{equation}
The likelihood $\mathcal{L}$ is then related to the sum over all spatial bins, over four logarithmic mass bins, and over the energy bands, as
\begin{equation}
\label{eq:Likelihood}\ln\mathcal{L} \simeq -\frac{1}{2}\sum_{\epsilon,\mytautheta,\Mbin}\chi^2(\epsilon,\mytautheta,\Mbin) \fin
\end{equation}
The test statistics TS \cite{MattoxEtAl96_TS}, defined as
\begin{equation}
\myTS \equiv -2\ln\frac{\mathcal{L}_{max,-}}{\mathcal{L}_{max,+}}\simeq\chi^2_{-} - \chi^2_{+} \coma
\end{equation}
can now be computed.
Here, subscript $-$ (subscript $+$) refers to the likelihood without (with) the modeled signal, maximized over any free parameters.
We define our nominal significance according to this TS test\RefMarkB{, assuming it follows a $\chi^2$ distribution with the number of degrees of freedom equal to the number of fit free parameters, without additional trial corrections}. The best-fit values and one-parameter confidence intervals are listed in Table \ref{tab:fit_res}.
The robustness of our results is demonstrated by several tests, some of which are detailed below.

\begin{deluxetable}{cc|ccccc|ccc}
\rotate
	\tablecaption{\label{tab:fit_res}
		\RefDelB{Model and }Parameter fit \RefMarkB{results}.
	}
    \tablewidth{1.3\textwidth}
	\tablehead{
	 $\tau$ range & $n_M$ & $\xi_{e}\dot{m}$ & $\mytaur_{s}$ & $s$ & $\myLAGN$ & $\mySagn$ & $\chi^2$ (dof) & $\myTS_{\mbox{\tiny ring}} (\sigma)$ & $\myTS_{\mbox{\tiny AGN}} (\sigma)$ \\
	 	(1) & (2) & (3) & (4) & (5) & (6) & (7) & (8) & (9) & (10)	}
	\startdata
    $\mathbf{[2.0,2.5]}$ & $\mathbf{4}$ & $0.8\pm0.2$ & $ \mathbf{2.4}$ & $\mathbf{-2}$ & $\mathbf{0}$ & {\bf--} & 11.9 (16) & $19.7\, (4.4\sigma)$ & ---\\
    $\mathbf{[2.0,2.5]}$ & --- & $0.81^{+0.02}_{-0.04}$ & $ \mathbf{2.4}$ & $\mathbf{-2}$ & $\mathbf{0}$ & {\bf--} & 427.4 (448) & $92.5 \,(9.6\sigma)$ & ---\\
    $\mathbf{[1.5,15]}$ & $\mathbf{4}$ & $0.5\pm0.1$ & $ 2.4\pm0.1$ & $\mathbf{-2}$ & $\mathbf{0}$ & {\bf--} & 350.5 (432) & $25.3\, (4.7\sigma)$  & --- \\
	$\mathbf{[1.5,15]}$ & --- & $0.34\pm0.06$ & $ 2.5\pm0.1$ & $\mathbf{-2}$ & $\mathbf{0}$ & {\bf--} & 10888 (12096) & $34.0 \,(5.5\sigma)$ & ---\\
	$\mathbf{[1.5,15]}$ & $\mathbf{4}$ & $1.9^{+40}_{-1.5} $ & $ 2.4\pm0.1$ & $-2.1\pm0.2$ & $\mathbf{0}$ & {\bf--} & 351.0 (432) & $24.8\, (4.6\sigma)$ & ---\\
	\hline
	$\mathbf{[0,15]}$ & $\mathbf{4}$    & $0.6\pm0.1$ & $ 2.3\pm0.1$ & $\mathbf{-2}$ & $16\pm3$ & $-1.5\pm0.2$	& 428.4 (480)&  $37.9\, (5.8\sigma)$& $38.1 \,(5.8\sigma)$ \\
	$\mathbf{[0,15]}$ & ---    & $0.54\pm0.05$ & $ 2.2\pm0.1$ & $\mathbf{-2}$ & $34\pm3$ & $-1.7^{+0.08}_{-0.06}$ & 12347 (13440) & $126.6\,(11.0\sigma)$& $236.4\,(15.2\sigma)$
		\enddata
	\tablecomments{Parameters in {\bf boldface} are constraints, rather than a result of the fit.
	\RefMarkB{Upper rows (above the horizontal line) show fits applied to only part of the data; the two bottom rows are our nominal fits for the full $\mytautheta$ range.}
\textbf{Columns:} (1) Range of normalized (to $R_{500}$) radial bins $\tau$ used in a fit; (2) \RefMarkB{Number of} logarithmic mass bins used; no mass bins are used for single-cluster co-addition; (3) Dimensionless CRE acceleration rate, in $1\%$ units; (4) Normalized (to $R_{500}$) shock radius; (5) Spectral index of ring emission photons; (6) AGN Luminosity in the emitted $(1\till100)\GeV$ energy range, in $10^{40}\erg\se^{-1}$ units; (7) Spectral index of the AGN signal; (8) $\chi^2$ value of the fit \RefMarkB{(and the number of degrees of freedom, before subtracting the number of free parameters, in parenthesis);} (9) TS value of adding the shell model signal (and the equivalent significance value in parenthesis; 1--3 free parameters); (10) TS of adding the AGN model signal (and the equivalent significance values in parenthesis; two free parameters).
	}
\end{deluxetable}

Fitting a central point-source model indicates the presence of such a source at the $5.8\sigma$ confidence level \RefMarkB{ $(\myTS=38.1)$}, based on the TS statistics. The fit suggests that the average cluster harbors a faint source, with a mean luminosity $\myLAGN=(1.6\pm0.3)\times10^{41} \erg \se^{-1}$ and a hard, $\mySagn=-1.5\pm0.2$ photon spectral index, consistent with AGN \gama-ray emission, in particular high synchrotron peak (HSP) BL Lacs \citep{Fermi09_TeVAGN, Fermi15_AGNcatalog}.
Indeed, four out of the five clusters with the most significant, $\nu_\sigma(\bm{\epsilon})>4$ signal in the central bin are known to harbor an AGN in their center: A3880 \cite{Sun09}, A3112, A3581 \cite{MittalEtAl09}, and A3744 \cite{WorrallBirkinshaw14}, where A3112 was already tentatively identified as a hard spectrum blazar \citep{AckermannEtAl14_GammaRayLimits}; the fifth cluster, RXC J2104.9-5149, is not well studied.
Based on the central flux and significance distributions among the different clusters in our sample, this signal can be crudely interpreted as one out of every four clusters in our sample harboring a point source of luminosity $\myLAGN\sim 7\times10^{41} \erg \se^{-1}$ in the emitted $(1\till100)\GeV$ band.
We note that such sources are too faint to be detected individually by the LAT, as the 3FGL catalog \cite{FermiPSC} detection limit is more than an order of magnitude brighter than the mean sources in our sample.
These conclusions support and extend previous claims for a faint population of \gama-ray AGN \cite{ProkhorovChurazov14, BranchiniEtAl17}.

\begin{figure}[t]
	\centerline{
		\epsfxsize=11cm \epsfbox{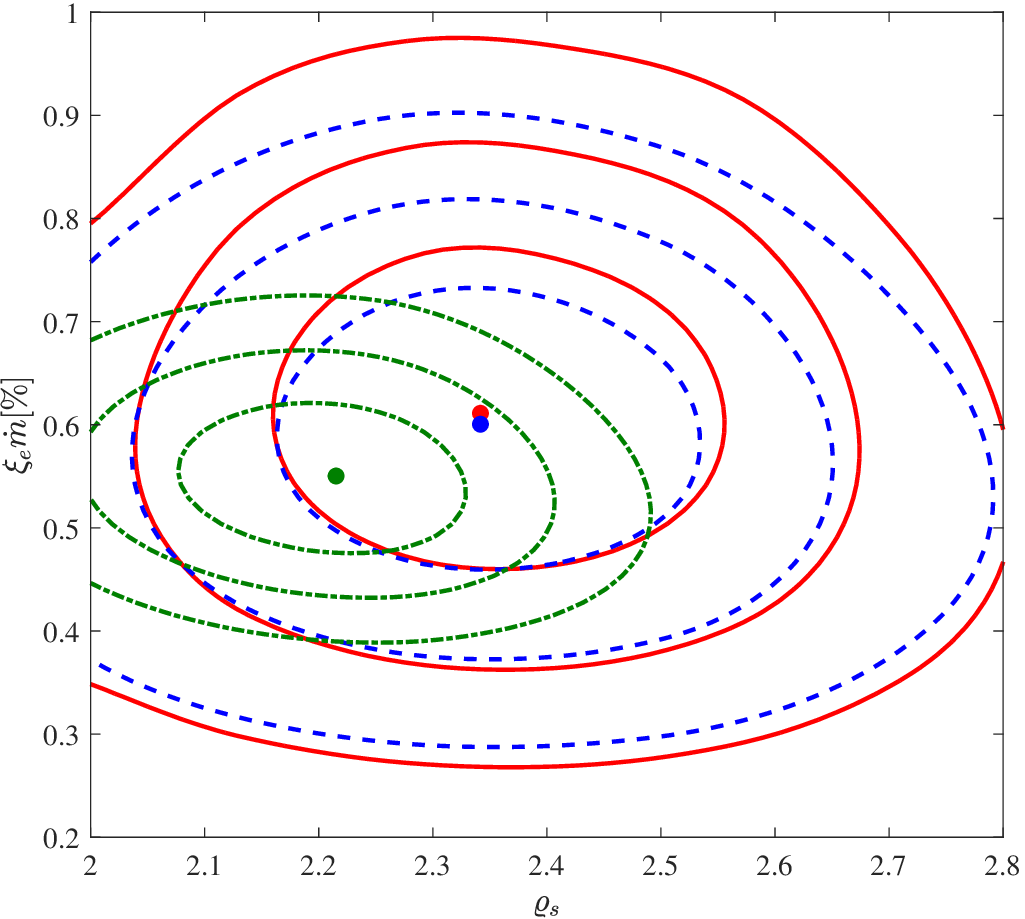}
	}
	\caption{\label{fig:chi2A}
		{\bf Confidence limits of the model parameters.}
		Two-parameter, $1\sigma$ through $3\sigma$ confidence intervals of the leptonic virial shell model parameters: CRE injection rate $\xi_e\dot{m}$, and scaled shock radius $\mytaur_{s}$. Shown is the two-dimensional cut through the best fit (disks) of the combined, ring and point source, four-parameter model. The results are shown for the nominal cluster $\beta$-models (red; solid contours), for modeling clusters as isothermal spheres in hydrostatic equilibrium (blue; dashed contours), and for a joint likelihood analysis (green; dot-dashed contours) .
	}
\end{figure}

Fitting a virial shell model (based on cluster $\beta$-models) indicates the presence of such a ring at a TS-based $5.8\sigma$ confidence level\RefMarkB{ $(\myTS=37.9)$}.
The fit suggests (Figure \ref{fig:chi2A}) CREs injected at a mean scaled shock radius $\mytaur_s=2.3\pm0.1$, at a rate $\xi_e\dot{m}=(0.6\pm0.1)\%$.
The signal is consistent with a flat, $s=-2$ photon spectral index, and changes little when modeling, masking, or removing the central sources; a ring-only model gives $s=-2.1\pm0.2$.
The calibrated model is consistent with the data in all four energy bands and in the four equal logarithmic mass bins used in the fit.
In the $\beta$-models of our sample, the mean dimensionless accretion rate is $\left<\dot{m}\right>\simeq4.2$.
If we adopt, instead, isothermal sphere models at hydrostatic equilibrium, $\xi_e$ nearly doubles, but $\dot{m}$ diminishes by a similar factor, leaving $\xi_e\dot{m}$ nearly unchanged (see Figure \ref{fig:chi2A}).
Our results are consistent with previous upper limits and with the signals in Coma (which was excluded from our sample).

\begin{figure}[t]
	\centerline{
		\epsfxsize=14cm \epsfbox{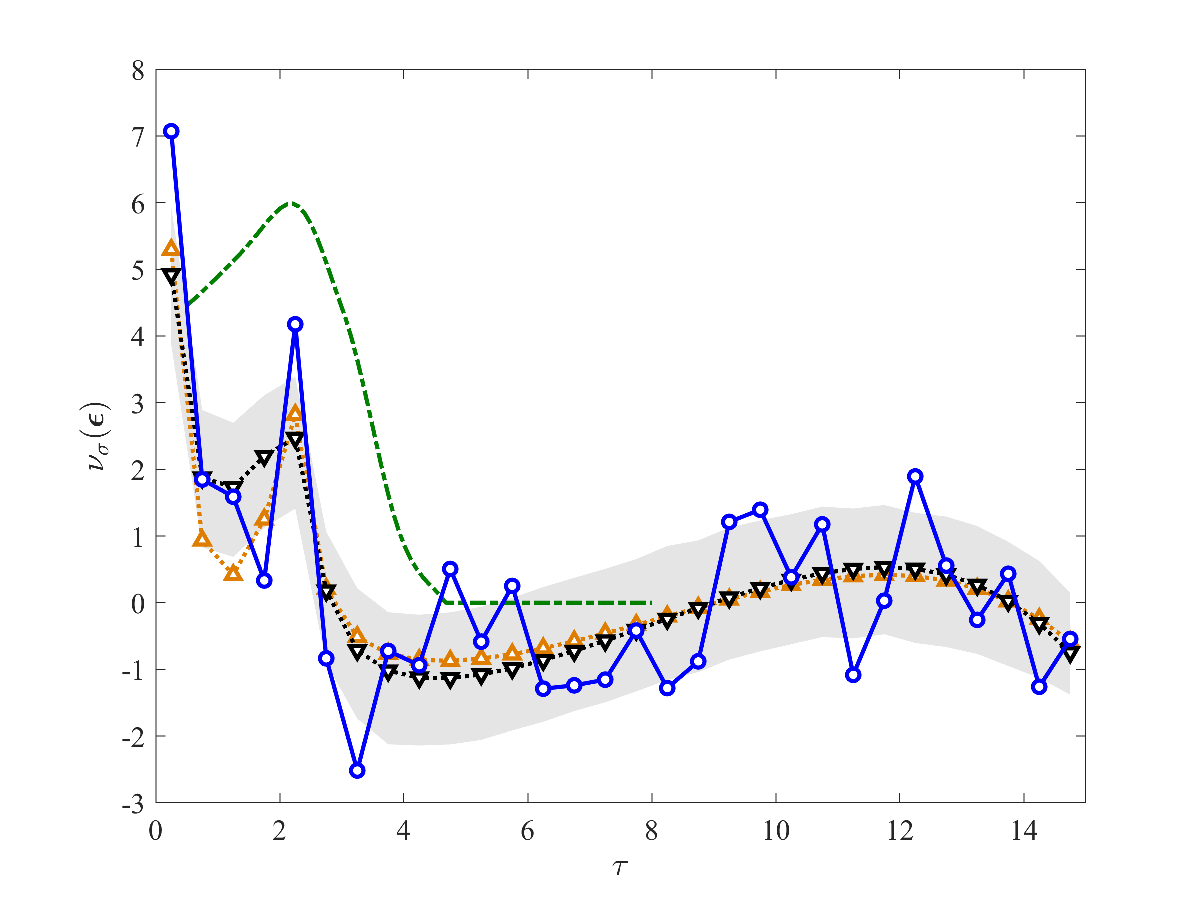}
	}
	\caption{\label{fig:combined2}
		\RefMarkA{{\bf Significance of the modeled signal.}
		The significance $\nu_\sigma(\bm{\epsilon})$ of the energy co-added excess \gama-ray counts over the foreground, shown as a function of $\mytautheta$ for the photon co-addition method (blue circles with solid lines to guide the eye).
		Also shown are the simulated signals for the best fit models combining AGN with a spherical virial shock ($\xi_{e}\dot{m}=0.6\%, \myLAGN=1.6\times10^{41} \erg \se^{-1}$; black down-triangles with a dotted line and with the $1\sigma$ extent of the mock catalog distribution as a shaded region, and with a planar shock (orange up-triangles with a dotted line).
		The TS-equivalent significance values of the full leptonic ring are also shown (green dash-dotted line).
	}}
\end{figure}

There is considerable freedom in the method in which the sample is co-added for the purpose of estimating the parameters and the signal significance.
This can be done by varying the mass bins used in Eqs.~(\ref{eq:ChiSquared}) and (\ref{eq:Likelihood}).
Our conservative, nominal mass binning uses four logarithmically spaced bins in the $10^{13}M_\odot$--$10^{15}M_\odot$ mass range; this is the binning method used to compute the above results.
The model resulting from this nominal mass binning is shown in \RefMarkA{Figure \ref{fig:combined2}} as black down triangles.

A different, extreme choice is to omit mass binning altogether, and replace the sum over mass bins in Eq.~(\ref{eq:Likelihood}) by a sum over clusters. This approach, analogous to the cluster co-addition of Eq.~(\ref{eq:clusterCoAddition}), and similar to a joint likelihood analysis \cite{AndersonEtAl15}, gives for the ring signal a very high TS \RefMarkB{value}, corresponding to a $11.0\sigma$ confidence level \RefMarkB{ $(\myTS=126.6)$}.
This enhanced significance reflects correlations between the signal and the model on a cluster-by-cluster basis.
The resulting model parameters, $\xi_e\dot{m}=(0.54\pm0.05)\%$ and $\mytaur_s=2.2\pm0.1$, are consistent with the nominal, more conservative mass binning results.
For the central signal we get an even higher TS, corresponding to a $15.2\sigma$ confidence level \RefMarkB{ $(\myTS=236.4)$}, with a hard, $\mySagn=-1.7^{+0.08}_{-0.06}$ spectrum consistent with the binned mass results.
The inferred source luminosity, $\myLAGN=(3.4\pm0.3)\times10^{41} \erg \se^{-1}$, is stronger here than with mass binning, probably due to a population of faint AGN that surface when the redshift of each cluster is incorporated in the model separately.

\RefDelB{
We test for possible effects of the central sources on the significance of the peripheral, ring signal, in two methods.
In the first method, we mask an increasingly larger inner region (only) when computing the TS.
When masking the central, $\mytautheta<1.5$ parts of the clusters instead of modeling the central source, the TS-based significance declines with respect to the nominal $5.8\sigma$, as expected from the omission of data points, but remains $>4.5\sigma$; masking a smaller region ($\mytautheta<1$) gives an intermediate, $>4.8\sigma$ significance.
The (ring) parameters evaluated in this method change by $<10\%$ with respect to their nominal values.
These results indicate that the central sources do not dominate the peripheral signal.
As a second, independent test, we repeat the analysis after removing clusters that show a significant central source, selected as $\nu_{\sigma,c}(\bm{\epsilon},\mytautheta<0.5)>2\sigma$.
This leaves a clean sub-sample of $93$ clusters, with no stacked central excess, $\nu_\sigma(\bm{\epsilon},\mytautheta<0.5)\simeq0$, yet showing a high, $5.7\sigma$ TS-based significance for the ring. In this test, $\mytaur_s$ changes by $<10\%$, but $\xi_e$ declines by $\sim 35\%$, partly because bright rings that contribute to the central bin were preferentially omitted.
}

To test if the ring signal is narrower (in $\mytautheta$) than the model, as may apparently (but not significantly, according to the $\chi^2$ values) seem from Figure \ref{fig:combined2}, we test if the stacking may have preferentially picked up shocks with brighter emission in the plane of the sky (as inferred in the Coma cluster \citep{KeshetReiss17}).
This model yields a signature (shown in the figure as orange up triangles) of width comparable to the signal, but of \RefDelB{nominal }\RefMarkB{lower} significance (\RefMarkB{$\myTS=20.6$, } $4.2\sigma$)\RefDelB{ lower} than that of the spherical shock model, and is therefore disfavored.
\RefMarkB{More sensitivity tests are provided in Appendix \ref{sec:sensitivity}.}

\section{Summary and discussion}
\label{sec:Discussion}

By stacking the $\sim8$ year, $(1-100)\GeV$ \emph{Fermi}-LAT data around the 112 most massive, extended, high-latitude galaxy clusters (see Figure \ref{fig:cluster mapA} and Table \ref{tab:clusters}), and radially binning the data, we find direct evidence for excess \gama-ray emission (see Figures \ref{fig:flux} and \ref{fig:combined1}) from these clusters.
Unlike previous studies, we rescaled the clusters to their angular radius $\theta_{500}$ before stacking them, and co-added the energy bands, utilizing the nearly flat energy spectrum.

We find two spatially-separated components, each arising from the cumulative contribution of many clusters.
Both components are marginally discernible in the folded, stacked image, even without radial binning (Figure \ref{fig:sig_mapA}).
To analyze these two components, we present a model for each one, weigh the measured signal against the simulated LAT signature in the model (taking into account PSF, foreground, binning, and stacking effects), and calibrate the model parameters using mock cluster catalogs.
The results are summarized in Table \ref{tab:fit_res}.

The central excess emission is unresolved, confined to the inner $0.5R_{500}$, with a significance of $\nu_\sigma(\bm{\epsilon})\simeq6\sigma\till7\sigma$.
This signal is morphologically consistent with a point source located at the center of each cluster.
Calibrating a point source model (see \S\ref{sec:AGNmodel}) indicates that such sources have a hard, $\mySagn=-1.5\pm0.2$ photon spectrum.
These properties suggest that the central signal is dominated by hard point sources such as AGN.
Indeed, the clusters most significantly contributing to the stacked central signal harbor a previously detected AGN or a tentative blazar.
A rough interpretation, based on the central flux and significance distributions among the different clusters in our sample, is that one out of every four clusters in our sample harbors a point source of luminosity $\myLAGN\sim 7\times10^{41} \erg \se^{-1}$ in the emitted $(1\till100)\GeV$ band, supporting and extending previous claims for a faint population of \gama-ray AGN \citep{ProkhorovChurazov14, BranchiniEtAl17}.
The significance of the point source model, evaluated using the test statistics TS, is $\sim 5.8\sigma$ \RefMarkB{ $(\myTS=38.1)$} in our nominal, conservative estimate, in which the clusters are divided into mass bins.
This significance rises to $15.2\sigma$ \RefMarkB{$(\myTS=236.4)$} if mass binning is relaxed, with little change in the inferred parameters.

The peripheral, ring-like excess peaks at $(2.0\till2.5)R_{500}$, with a significance of $\nu_\sigma(\bm{\epsilon})\simeq4.2\sigma$. This signal matches the expected signature of \gama-ray rings arising from inverse-Compton scattering of CMB photons by virial-shock accelerated CREs.
A \gama-ray shell model (based on $\beta$-models or isothermal sphere models of the clusters; \S\ref{sec:RingModel}) suggests (Figure \ref{fig:chi2A} and Figure \ref{fig:combined2}) CREs accelerated on average at a $\mytaur_s=2.3\pm0.1$ scaled shock radius, at a dimensionless acceleration rate $\xi_e \dot{m}=(0.6\pm0.1)\%$.
The model is consistent with the anticipated flat, $s=-2$ photon spectrum, and changes little when masking the central source.
These calibrated models are consistent with the data in all four energy bands and in all four mass bins. The significance of the modeled ring signal, evaluated using the test statistics TS, is $\sim 5.8\sigma$ \RefMarkB{ $(\myTS=37.9)$} in our nominal, mass-binned method.
This confidence level rises to $11.0\sigma$ \RefMarkB{ $(\myTS=126.6)$} if mass binning is relaxed, with no significant change in the inferred parameters.
These results provide a high-significance detection of the virial shock, and reveal correlations between model and signal on a cluster-by-cluster basis.

\RefMarkB{
The above results are obtained in a method purposely chosen to be conservative, and shown to be robust.
We repeat the analysis twice (see the two bottom rows in Table \ref{tab:fit_res}), with and without mass binning. Both variants detect both signals (central and peripheral) at a high confidence level.
We therefore adopt the more conservative --- mass binned --- of the two methods, and use its results as our nominal significance values.
A suite of convergence and sensitivity tests (see Appendix
\ref{sec:sensitivity}) indicate that our results are robust to variations in the preparation of the LAT data and of the cluster sample, in the photon analysis methods, in the cluster stacking methods, and in our energy co-addition method, and we found only a small effect of the central source on the ring signal.
For example, we show that the low-significance, dip-like effect seen just outside the peripheral signal in Figures \ref{fig:combined1} and \ref{fig:combined2} is associated with two regions on the sky in which the virial radii of multiple clusters overlap; excluding these overlapping clusters completely removes the dip, and slightly enhances the significance of the peripheral signal.
}

Our results are consistent with previous upper limits and with the broad band signal in the Coma cluster (which was excluded from our sample).
In particular, our stacked estimate $\xi_e\dot{m} \simeq0.6\%$ is similar to the $\xi_e\dot{m}\simeq0.3\%$ value inferred from the elliptical soft X-ray and \gama-ray ring in Coma \citep{KeshetReiss17}.
Both estimates are well below previous upper limits, \eg $\xi_e<1\%$ in Coma \citep{ZandanelAndo14}.
The best previous upper limit \cite{GriffinEtAl14} on the integrated \gama-ray photon flux from a galaxy cluster was $2.3\times10^{-11} \se^{-1} \cm^{-2}$, at a $95\%$ confidence level in the energy range $(0.8\till100)\GeV$ for clusters of typical mass $M_{200}\sim 6\times 10^{14}M_\odot$.
This is similar to our corresponding measurement, $\sim2.8\times 10^{-11} \se^{-1}\cm^{-2}$ for the entire ($<2.5R_{500}$) emission from a typical cluster in our sample.

Our analysis is able to reach sensitivities significantly better than previous stacking analyses due to a novel combination of angular rescaling prior to stacking, radial binning, reliance on high energies where the PSF is small, and energy band and mass-bin co-addition, as well as a somewhat longer LAT integration time.
\RefDelA{Our TS estimates based on cluster and energy co-addition resemble image co-addition methods argued to be optimal \citep{ZackayOfek17}.}

The stacked ring signal we find does not necessarily imply that any individual cluster is surrounded in projection by a circular \gama-ray ring.
Indeed, individual cluster rings were predicted to be patchy, and the same applies for the stacked ring at present significance levels.
\RefMarkC{Numerical studies suggest that there is substantial variability in the position, morphology, and energy dissipation properties, among the shocks of different clusters (\eg \cite{KeshetEtAl03,KeshetEtAl04, VazzaEtAl09, Miniati14, SchaalSpringel15}).}
The LAT data were radially binned (and in Figure \ref{fig:sig_mapA}, rotated randomly around the center of each cluster), so the ring signature reflects the averaging over various \gama-ray morphologies, including elliptic, asymmetric, and effectively spatially-intermittent patterns.
Our estimate of the scaled shock radius $\mytaur_s$ is the mean projected radius of these \gama-ray morphologies.
\RefMarkC{
While simulated shock features can be seen in a wide range of radii (\eg \cite{KeshetEtAl03, VazzaEtAl09, Miniati14, SchaalSpringel15}), the relevant scale for the present analysis is the mean radius in which the dissipation of energy by strong shocks into \gama-ray emitting CREs is maximized, when averaged over many clusters.
Estimates of this scale lie near the virial radius \citep{KeshetEtAl03, KeshetEtAl04}. This scale is closely related to the radius in which the mean entropy injection rate is maximal (when averaged over many clusters), $\sim1.3R_{200}\simeq 2.1R_{500}$ \citep{SchaalSpringel15}.
This is indeed comparable to the typical shock radius inferred from simple spherical collapse models \cite{EkeEtAl96}, and is close to the radius of our \gama-ray ring.
It should be noted, however, that such simulations and simplified models incorporate only part of the relevant physics, often neglecting, for example, radiative cooling, photoionization of the intergalactic medium, and the effects of the accelerated particles themselves.
}

The acceleration rate we measure is similarly a global average of the local CRE injection at this (non-projected) scaled radius.
Therefore, our best fit $\xi_e\dot{m}\sim 0.6\%$, with an uncertainty factor of $\sim 2$ dominated by interpretation systematics (in cluster thermal models, deviations from equilibrium and spherical symmetry, and virial shock modeling), may underestimate the true acceleration rate.

Adopting our nominal CRE injection rate as typical of all clusters, we obtain \cite{KeshetEtAl04} a diffuse \gama-ray component $\epsilon^2 dJ/d\epsilon\simeq 40(\xi_{e}\dot{m}/1\%)\eV\se^{-1}\cm^{-2}\sr^{-1}$,
contributing a significant fraction (\eg $15\%$ at $100\GeV$) of the high-energy extragalactic \gama-ray background \cite{AckermannEtAl12}.
In the radio, we find a $\nu I_\nu \sim 10^{-11}(\xi_{e}\dot{m}/1\%) (\xi_B/1\%)\erg\se^{-1}\cm^{-2}\sr^{-1}$ synchrotron signal,
where $\nu$ is the frequency and $I$ is the intensity.
This signal is observable through $\delta T_l\simeq 0.4 (\nu/\mbox{GHz})^{-3}\K$ fluctuations at multipoles $400< l < 2000$ with present interferometers such as LOFAR and EVLA \cite{KeshetEtAl04, KeshetEtAl04_SKA}.

Lately, an analysis of the \emph{Planck} $y$-parameter map of A2319 \cite{HurierEtAl17} yielded the first high-significance detection ($8.6\sigma$) of the anticipated drop in the SZ pressure near the virial radius of a galaxy cluster, indicating a strong shock with Mach number $\Upsilon>3.25$ (at $95\%$ confidence level), located at $(2.93\pm0.05)\myR$.
Comparing SZ and \gama-ray maps \cite{KeshetEtAl18} indicated similar drops in $y$-parameter also in Coma ($4.1\sigma$) and in A2142 ($3.1\sigma$), consistent with strong shocks, and coincident with a LAT \gama-ray excess. The acceleration rates $ \xi_e \dot{m}$ inferred in these three clusters \cite{KeshetEtAl18} are of order a few $0.1\%$, similar to the present stacking analysis.

\RefMarkB{
	We examine if there are any correlations between the \gama-ray emission --- either from the center or from the periphery --- and various attributes of the clusters, such as their X-ray properties.
	Such correlations are less sensitive than our TS analysis, mainly due to the reduced number of degrees of freedom. Indeed, although some correlations can be pointed out, we do not find a significant result, neither for individual clusters, nor for a coarsely-binned population. For example, we test for correlations with the dynamical state of the
	cluster (\ie cool-core, relaxed clusters vs. non-cool-core, merging clusters). Here, in addition to the low number of degrees of freedom, \myNi there are different criteria for classifying the dynamical state; \myNii such classifications are available for only a small subset of the MCXC clusters; and \myNiii the classified clusters are biased toward the high mass end.
	Consequently, we do not find any significant, unbiased trend.
}

Our detection of \gama-ray rings around clusters confirms the paradigm of LSS accretion through virial shocks.
Stacking shows a signal in spite of the dispersion in the radii and morphologies of the shocks, suggesting a substantial population of shocks that are not highly non-spherical in projection.
The signal is consistent with shocks lying at a nearly fixed enclosed over-density $\delta$, as accordingly rescaling their radius has facilitated the detection of the stacked signal; furthermore, the shock location closely matches that expected from simple spherical collapse models \cite{EkeEtAl96}, and from simulated $\Lambda$CDM clusters \cite{KeshetEtAl03, SchaalSpringel15}.
Our results positively test the theory of CRE acceleration, generalizing it to scales much larger than accessible ever before.
Resolving individual shocks in the future will teach us much about LSS, and in particular its growth.

\appendix
\section{Emission models}
\label{sec:BetaModel}

An analytic model for the \gama-ray emission from a galaxy cluster requires some assumptions specifying the gas distribution.
Simple choices include an isothermal sphere \cite{WaxmanLoeb00, KeshetEtAl04} or an isothermal $\beta$-model \cite{KushnirWaxman09}.
We adopt the latter, as it underlies much of the MCXC catalog. Here, the number density of thermal electrons is given by
\begin{equation}\label{eq:nBeta}
n_e = n_0 \left[1+\left(\frac{r}{r_c}\right)^2\right]^{-3\beta/2} \coma
\end{equation}
where $n_0$ is the central electron number density, $r_c$ is the core radius, and $\beta$ is the slope parameter.
Note that the isothermal sphere distribution is a special case of the isothermal $\beta$ model, corresponding to $\beta=2/3$ in the $r_c\to0$ limit.

We assume that the cluster is approximately in hydrostatic equilibrium, implying that the total (gravitating) mass inside a radius $r$ is
\begin{equation}
M(r) \simeq \frac{3\beta k_BT r}{G \mass}\left(1+\frac{r_c^2}{r^2}\right)^{-1} \coma
\end{equation}
where $k_B$ is the Boltzmann constant, $G$ is Newton's constant, $\mass\simeq 0.59m_p$ is the mean particle mass, and $m_p$ is the proton mass.
Let $\delta=100\delta_{100}$ be the over-density parameter, defining a radius $R_\delta$ and an enclosed mass $M_\delta$, such that the mean enclosed mass density $M_\delta/[(4\pi/3)R_\delta^3]$ is higher by a factor of $\delta$ than the critical mass density $\rho_c$ of the Universe; we use the value of $\delta$ as a subscript for $r$, $M$, and $\delta$ itself.

Consider radial distances $r>\myR$ from the center of the cluster, where we may neglect the core (of typical radius $0.1R_{500}$, as found for our sample in Table \ref{tab:clusters}) and approximate $n_e\simeq n_0(r/r_c)^{-3\beta}$.
This leads to the approximate mass
\begin{equation}
\label{eq:M_T relation}
M(r)\simeq \frac{3\beta k_B T}{G\mass}r \simeq 3.8\times 10^{14}\myeta T_5 \frac{r}{1\Mpc} M_\odot \coma
\end{equation}
radius
\begin{equation}
R_\delta = \left(\frac{9\beta k_B T}{4\pi \delta G \mass \rho_c } \right)^{1/2} \simeq 2.6 \left( \frac{\myeta T_5}{\delta_{100}\myEz} \right)^{1/2} \Mpc \, ,
\end{equation}
and mass--temperature relation
\begin{equation} \label{eq:MvsT}
M_\delta = \frac{9/2}{\sqrt{\pi \rho_c \delta}} \left(\frac{\beta k_B T}{G\mass} \right)^{3/2} \simeq 10^{15} (\eta T_5)^{3/2}\delta_{100}^{-1/2} \myEz^{-1} M_\odot \coma
\end{equation}
where $T_5\equiv k_BT/5\keV$, we defined $\myEz(z)\equiv H(z)/H_0 \simeq [(1-\Omega_m)+(1+z)^3\Omega_m]^{1/2}$ as the ratio between the Hubble constant at redshift $z$ and its present day value, $\Omega_m$ is the matter fraction of the Universe, and $\myeta\equiv\beta/(2/3)$ is the normalized profile index, which becomes unity for the standard, isothermal sphere slope.

\subsection{Leptonic $\gamma$-ray shell model}
\label{sec:RingModel}

In the strong shock limit, the downstream velocity with respect to the shock is
\begin{equation}\label{eq:vdStrongShock}
v_d = \left[\frac{(\Gamma-1)k_BT}{2\mass}\right]^{1/2} = \left(\frac{k_BT}{3\mass}\right)^{1/2} \coma
\end{equation}
where $\Gamma=5/3$ is the adiabatic index of the gas,
so we may compute the dimensionless accretion rate through $R_\delta$ as
\begin{equation}
\dot{m} \equiv \frac{\dot{M}}{MH} \simeq \frac{4\pi R_\delta^2 \bar{m} n_d v_d}{f_b M_\delta H_0 \myEz} \fin
\end{equation}
Here, we assumed that the baryon mass enclosed inside $R_\delta$ satisfies $M_{b,\delta}=f_b M_\delta$, where $f_b\simeq 0.17$ is the cosmic baryon fraction.

If the $\beta$-model parameters (including $n_0$ and $r_c$) are known, one may compute $n_d$ and evaluate $\dot{m}$.
In the absence of complete models for all clusters in the sample, here we estimate $\dot{m}$ without these parameters, by assuming that $M_b$ is given by the spatial integral of Eq.~(\ref{eq:nBeta}). This implies that
\begin{equation}
n_d = (1-\beta) \frac{f_b \rho_c}{\mass}\delta \simeq 5.3\times 10^{-5} (3-2\eta)\delta_{100} \myEz^2 \cm^{-3} \coma
\end{equation}
and so
\begin{equation} \label{eq:mdot1}
\dot{m}
\simeq (3-2\eta)H^{-1} \left(\frac{k_BT}{3\mass}\right)^{1/2} \left(\frac{4\pi \rho_c  \delta}{3M_\delta}\right)^{1/3} \simeq  4.7 \left(3-2\eta\right)\left(T_5 \delta_{100}\right)^{1/2}(\myEz M_{14})^{-1/3} \coma
\end{equation}
where $M_{14}\equiv M_{500}/10^{14}M_\odot$, and we used $M_\delta\propto \delta^{-1/2}\propto R_\delta$.
If we use the mass-temperature relation Eq.~(\ref{eq:MvsT}), implicitly invoking hydrostatic equilibrium again, this simplifies to
\begin{equation}
\dot{m} \simeq \frac{2(1-\beta)}{H} \sqrt{\frac{\pi \delta G\rho_c}{3\beta}} \simeq 2.9 \left( \frac{3-2\eta}{\eta^{1/2}} \right) \delta_{100}^{1/2} \fin
\end{equation}

We adopt the Fermi diffusive shock acceleration model, in which CREs are accelerated with a flat spectrum up to a CMB scattering-limited Lorentz factor
\begin{eqnarray}
\gamma_{max} & \simeq &  \frac{\Gamma+1}{2} \sqrt{\frac{3 e B k_B T}{(\Gamma-1)m_p c^2\sigma_T u_{cmb}}}  \simeq  1.7\times 10^8 \left[(3-2\eta)\xi_{B1}\delta_{100}\right]^{1/4} T_5^{3/4} \nonumber \\  & \simeq & 8.2 \times 10^7 \left[\frac{(3-2\eta)\xi_{B1}\delta_{100}}{\eta^3}\right]^{1/4} M_{14}^{1/2} \coma
\end{eqnarray}
where $c$ is the speed of light, $\sigma_T$ is the Thompson cross section, $u_{cmb}$ is the CMB energy density, $B$ is the magnetic field, we assumed that the magnetic energy downstream is a fraction $\xi_B\equiv 0.01\xi_{B1}$ of the thermal energy, and in the last line we used Eq.~(\ref{eq:MvsT}).
The inverse-Compton emissivity per unit shock area may now be computed for a flat CRE spectrum as
\begin{eqnarray} \label{eq:ICEmissivityBeta}
\epsilon^2  \frac{dN}{dt\, dA\,  d\epsilon} & \simeq & \frac{(3/2)\xi_e f_b \dot{M}_\delta k_B T}{(4\pi R_\delta^2) \mass (2\ln \gamma_{max})} \simeq \frac{3^{1/2}(1-\beta)}{4}\frac{f_b \xi_e \rho_c \delta}{\ln\gamma_{max}} \left(\frac{k_B T}{\mass}\right)^{3/2} \nonumber \\
& \simeq & 9.0 \times 10^{-9} \myEz^2 \left(3-2\eta\right)\xi_{e,1}\delta_{100} T_5^{3/2} \erg \se^{-1} \cm^{-2}  \fin
\end{eqnarray}
As the maximal CRE energy enters this expression only logarithmically, here and below we adopt $\gamma_{max}\simeq 10^8$, neglecting the logarithmic dependence on redshift and cluster properties.
The resulting photon energy flux from the entire, spherical shock may now be written as
\begin{eqnarray} \label{eq:ICFluxBeta}
\epsilon \frac{dF}{d\epsilon} & = & \frac{4\pi R_\delta^2}{4\pi d_L^2} \epsilon^2 \frac{dN}{dt\, dA\,  d\epsilon} =  \frac{\theta_\delta^2}{(1+z)^4} \epsilon^2\frac{d N}{dt\, dA\,  d\epsilon}   \\
& \simeq & 5.5\times 10^{-13} \xi_{e,1}  \frac{\left(3-2\eta\right)\myEz^2 T_5^{3/2} \theta_{0.2}^2}{(1+z)^4} \erg \se^{-1} \cm^{-2} \coma \nonumber
\end{eqnarray}
where $d_L$ is the luminosity diameter distance, and $\theta_{0.2}\equiv\theta_{500}/0.2\dgr$ is the normalized angular equivalent of $R_{500}$.
Notice that this result is independent of the actual radius of the shock.

Using Eq.~(\ref{eq:MvsT}), the emissivity may alternatively be written in terms of cluster mass, invoking the hydrostatic-equilibrium assumption to suppress the temperature dependence,
\begin{eqnarray} \label{eq:ICEmissivityBetaM}
\epsilon^2 \frac{dN}{dt\, dA\,  d\epsilon}
&\simeq & \frac{(1-\beta)}{6\ln\gamma_{max}} \sqrt{\frac{\pi}{3}}\left( \frac{G \rho_c \delta }{\beta}\right)^{3/2} f_b \xi_e M_\delta \\
& \simeq & 2.1 \times 10^{-9} \myEz^3 \left(\frac{3-2\eta}{\eta^{3/2}}\right) \xi_{e,1} \delta_{100} M_{14} \erg \se^{-1} \cm^{-2} \nonumber \coma
\end{eqnarray}
which yields the photon energy flux
\begin{eqnarray} \label{eq:ICFluxBetaM}
\epsilon \frac{dF}{d\epsilon}
& \simeq & 1.3\times 10^{-13} \xi_{e,1} \left(\frac{3-2\eta}{\eta^{3/2}}\right) \frac{ \myEz^3 M_{14} \theta_{0.2}^2}{(1+z)^4} \erg \se^{-1} \cm^{-2} \fin \nonumber \\
\end{eqnarray}

Another alternative is to relate the downstream velocity not to the temperature as in Eq.~(\ref{eq:vdStrongShock}), but rather to the cluster's mass, by assuming that the upstream gas has free-fallen from rest until crossing the shock at $R_\delta$,
such that
\begin{equation} \label{eq:Strange}
v_d \simeq \frac{\Gamma-1}{\Gamma+1} v_u \simeq \frac{\Gamma-1}{\Gamma+1} \left( \frac{2GM_\delta}{R_\delta} \right)^{1/2} \fin
\end{equation}
This gives rise to slightly different scalings of the signal,
\begin{equation} \label{eq:FreeFallFluxBeta}
\epsilon \frac{dF}{d\epsilon} \simeq 2.9\times 10^{-13} \xi_{e,1}  \frac{\myEz^{7/3}\left(3-2\eta\right)M_{14}^{1/3} T_5 \theta_{0.2}^2}{(1+z)^4}\erg \se^{-1} \cm^{-2} \fin
\end{equation}

We have varied our interpretation of the LAT data analysis by using either Eq.~(\ref{eq:ICFluxBeta}) or Eq.~(\ref{eq:ICFluxBetaM}) or Eq.~(\ref{eq:FreeFallFluxBeta}). The results change by less than $20\%$.

\subsection{AGN model}
\label{sec:AGNmodel}

The model we adopt for the \gama-ray emission from AGN assumes a point source with a power-law spectrum. Note that unlike the inverse-Compton emission from the CREs produced in the virial shock, AGN are known to show a wide variety of spectral indices \cite{Fermi15_AGNcatalog}, $-3.2<\mySagn<-1.2$.
The power-law spectrum gives rise to a received photon energy flux
\begin{equation}
\label{eq:AGNFlux}
\epsilon F_{\epsilon} = \frac{\myLAGN }{C}\frac{(1+z)^{\mySagn}}{4\pi d_L^2} \epsilon^{\mySagn+2} \coma
\end{equation}
Where $\myLAGN$ is the luminosity in the emitted $(1\till100)\GeV$ band, and
\begin{equation}
C=\int\limits_{1\GeV}^{100\GeV}\epsilon^{\mySagn+1}d\epsilon
\end{equation}
is the normalization constant.

\section{Sensitivity tests}
\label{sec:sensitivity}

\RefMarkA{
We perform various sensitivity and consistency tests to ensure that our estimates are robust and conservative.}

\RefMarkB{
\subsection{Analysis method tests}
}

\begin{figure}[t]
	\centerline{\epsfxsize=8.4cm \epsfbox{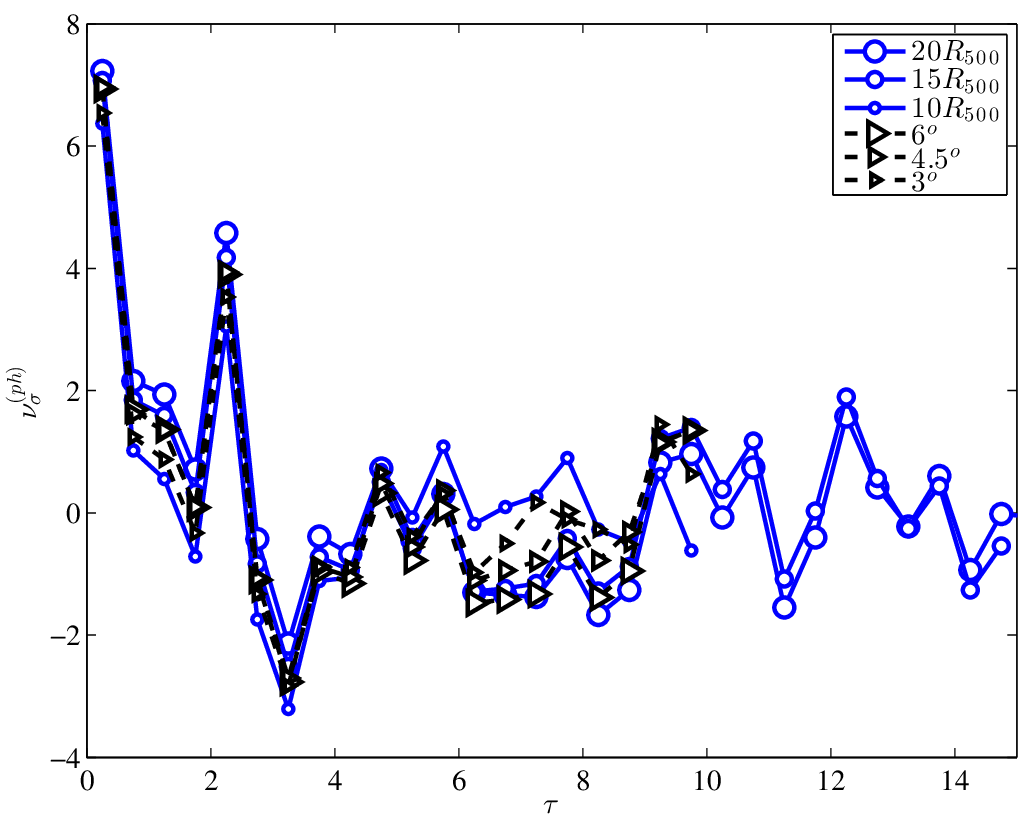}}
	\caption{\label{fig:diff Fg size} The photon and energy co-added significance of the excess counts as a function of $\mytautheta$ for different foreground estimation regions (see $\mytautheta_{max}$ and $\theta_{max}$ values in the legend).}
\end{figure}

\RefMarkA{
In Figure \ref{fig:diff Fg size} we check for the size effect of the area used to fit the foreground, both as a fixed scaled radius $[10\till20]\myR$, or as a constant angular distance in the range $3\dgr\till6\dgr$. We find no significant variations in $\nu_\sigma$.}

\begin{figure}[t]
	\centerline{\epsfxsize=8.4cm \epsfbox{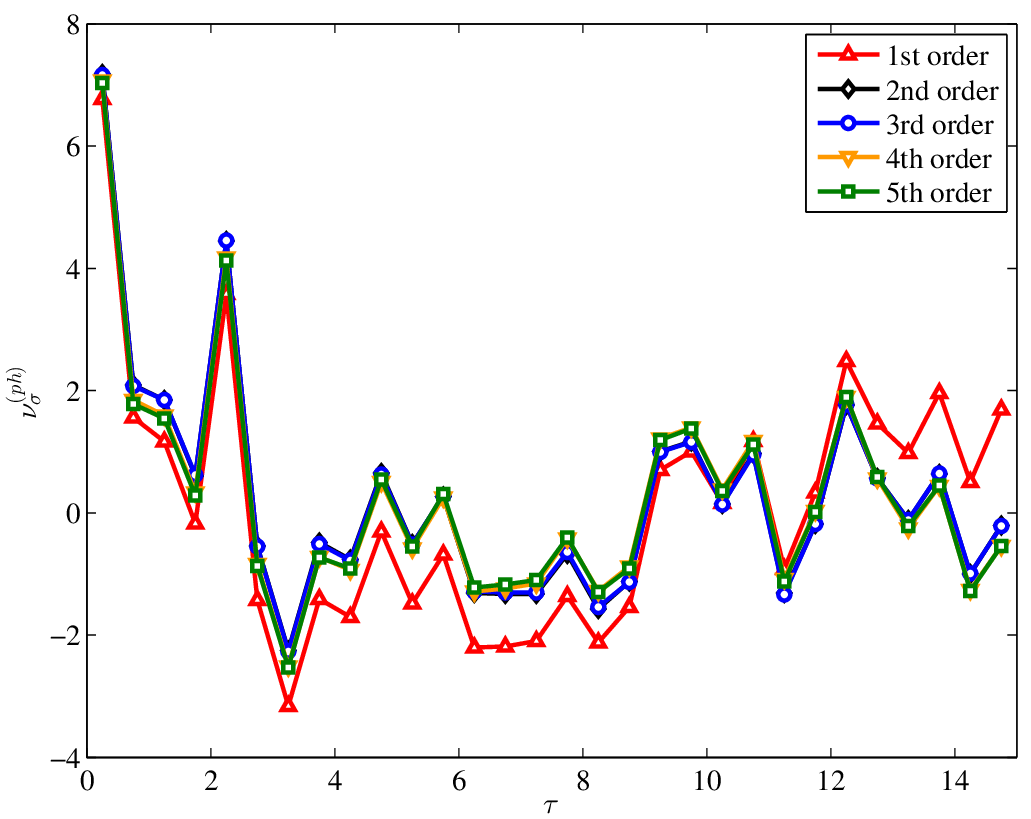}}
	\caption{\label{fig:diff Fg}
		The photon and energy co-added significance of the excess counts as a function of $\mytautheta$ for different polynomial order $N_f$ of foreground estimation (see legend).
	}
\end{figure}

\RefMarkA{
To test the nominal significance inferred from the foreground estimate, we examine different models for the foreground determination, ranging from 1st order to 5th order polynomials, as presented in Figure \ref{fig:diff Fg}. The results (in the $<3\myR$ region) change only little.}

\begin{figure}[t]
	\centerline{\epsfxsize=8.4cm \epsfbox{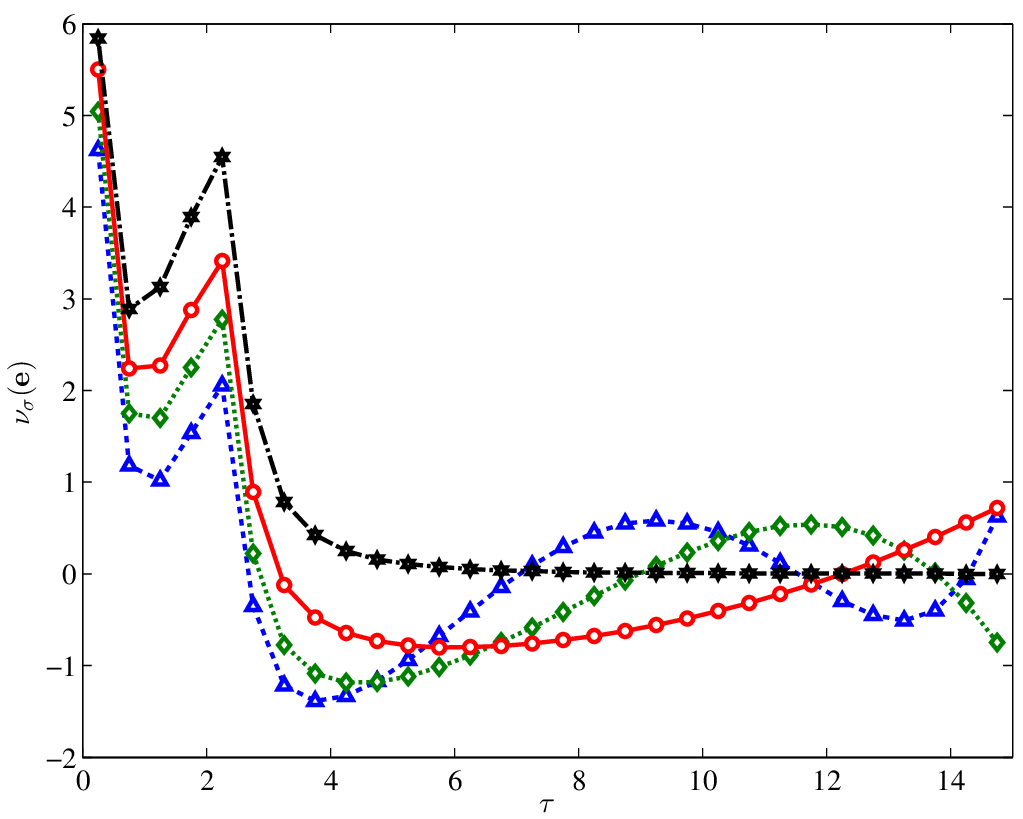}}
	\caption{\label{fig:S_wave} The photon and energy co-added significance of the excess counts of the mock catalog with the best fit model, as a function of $\mytautheta$, for different polynomial order $N_f$ of foreground estimation: $N_f=2$ (red circles with solid line to guide the eye), $N_f=4$ (green diamonds with dotted line to guide the eye), and $N_f=6$ (blue triangles with dashed line to guide the eye). Also show is the result for the actual mock foreground (black pentagrams with dash-dotted line to guide the eye).}
\end{figure}

\RefMarkA{
In Figure \ref{fig:S_wave} we show how changing the order $N_f$ of the foreground fit modifies the 'S' pattern that appears in a mock sample, once a simulated signal has been added to it (using our best fit values). It can be seen that, as expected, when the fit order increases, the wavelength of the wiggles shortens.}

\RefMarkA{
As $N_f$ increases, the significance one would attribute to the signal slowly decreases, as the foreground fit can follow the signal increasingly well, thus attributing part of the signal to the foreground. The TS value calculated from the fit is thus underestimated (compared to the true value using the true foreground, which is known in the mock catalog), by a factor of $\sim 2$ for fit orders $2\till 6$.}

\begin{figure}[t]
	\centerline{\epsfxsize=9.4cm \epsfbox{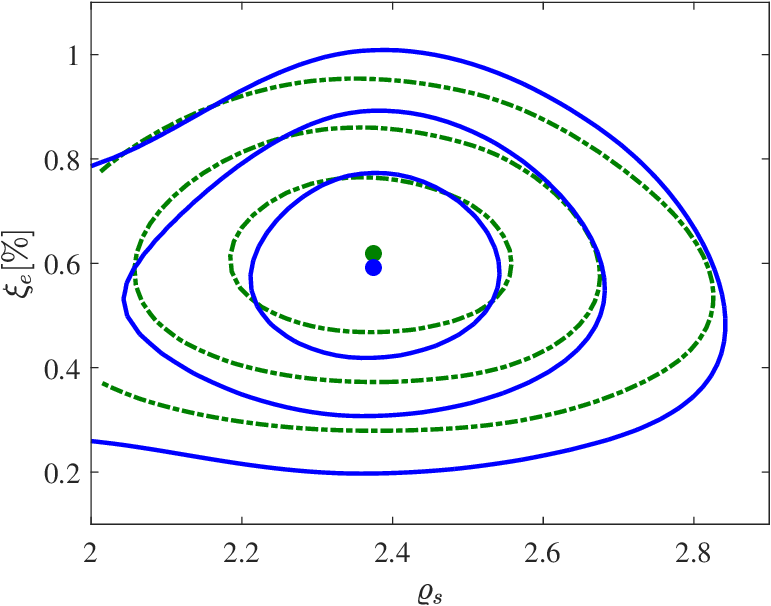}}
	\caption{\label{fig:4vs6} 	Two-parameter, $1\sigma\till3\sigma$ confidence intervals of the ring model parameters, CRE injection rate $\xi_e \dot{m}$ and scaled shock radius $\mytaur_v$. The $\chi^2$ values were calculated by co-adding the (photon co-added) four mass bins, where the foreground is fitted using $N_f=4$ (dashed red) or $N_f=6$ (solid green). Best fit values are shown as disks. The two best values are insignificantly different, while the confidence intervals are bigger for $N_f=6$
	}
\end{figure}

\RefMarkA{
In contrast, the model parameter values one would infer using the foreground are not sensitive to the details of the foreground determination.
In fact, using a $N_f=6$ gives best-fit parameters very similar to those of the nominal, $N_f=4$ ring model, albeit with larger confidence interval, as seen in Figure \ref{fig:4vs6}.}

\begin{figure}[t]
	\centerline{\epsfxsize=8.4cm \epsfbox{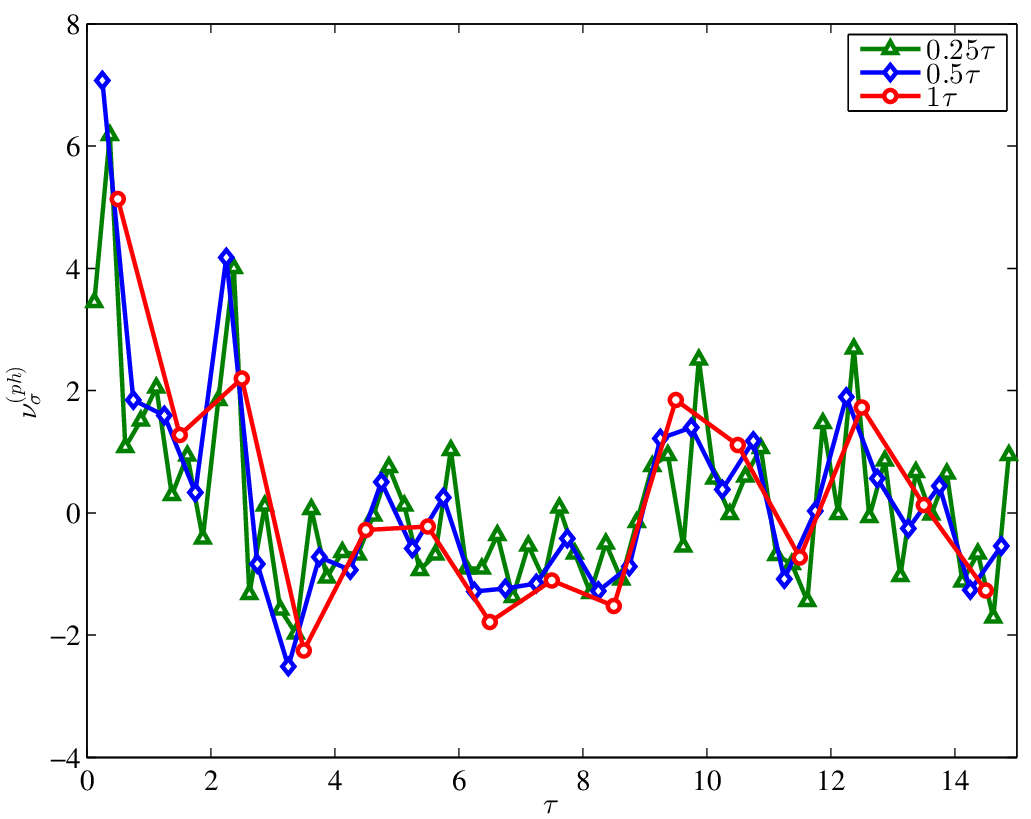}}
	\caption{\label{fig:bin size}
		The photon and energy co-added significance of the excess counts as a function of $\mytautheta$ for different choices of bin sizes $\Delta\mytautheta$ (see legend).
	}
\end{figure}

\RefMarkA{
In Figure \ref{fig:bin size} we examine the dependence of the results upon the radial bin size, $\Delta \mytautheta$. Different choices, namely $\{0.25, 0.5, 1\}\myR$, for the bin size are shown. It can be seen that the significance of the peak and the ring signals are only slightly affected by these variations.}

\begin{figure}[t]
	\centerline{\epsfxsize=8.0cm \epsfbox{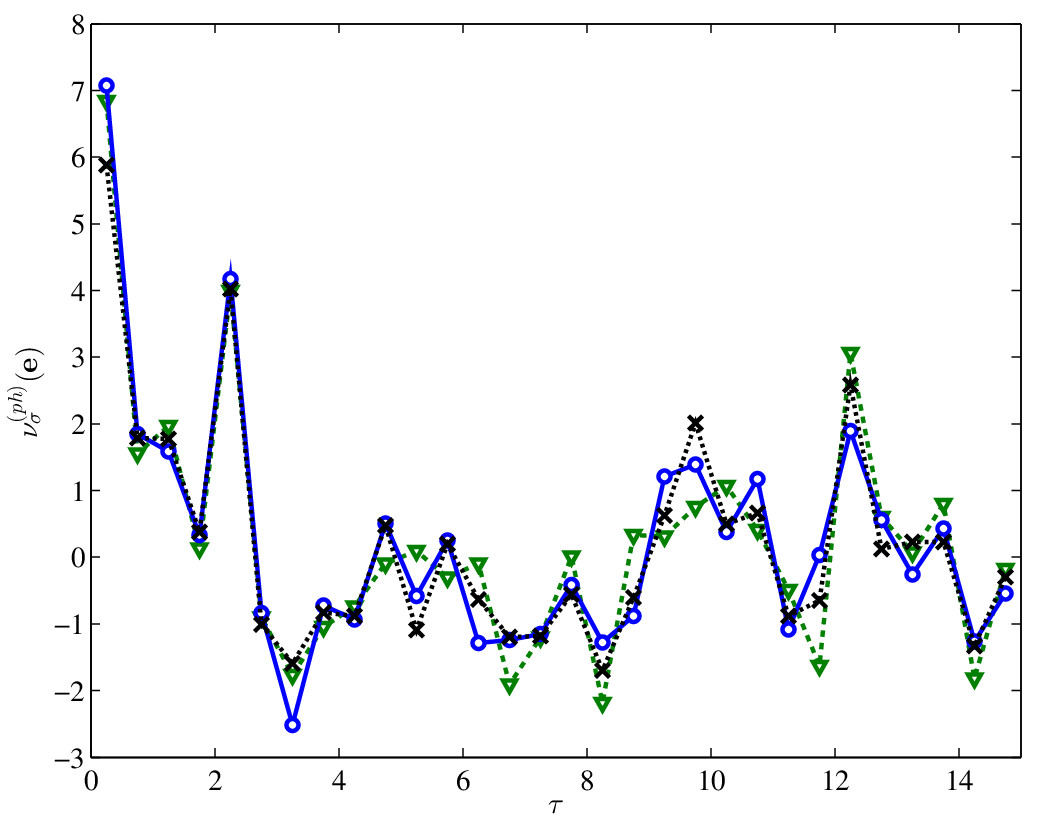}
	\epsfxsize=8.0cm \epsfbox{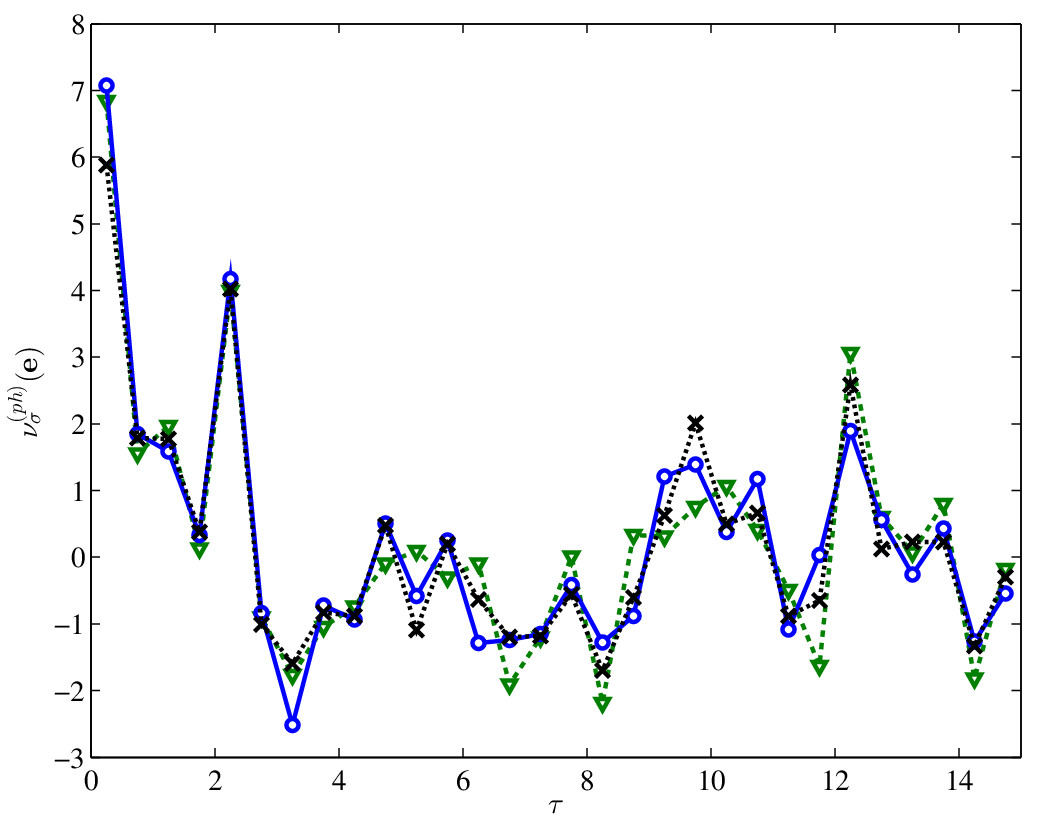}}
	\caption{\label{fig:healpix10}
		The photon co-added (left panel) or cluster co-added (right panel), energy co-added significance of the excess counts as a function of $\mytautheta$, using HEALPix order $N_{hp}=9$ (green triangles with dashed line to guide the eye), $N_{hp}=10$ (blue circles with solid line to guide the eye), and $N_{hp}=11$ (black crosses with dotted line to guide the eye).
	}
\end{figure}

\begin{figure}[t]
	\centerline{\epsfxsize=8.4cm \epsfbox{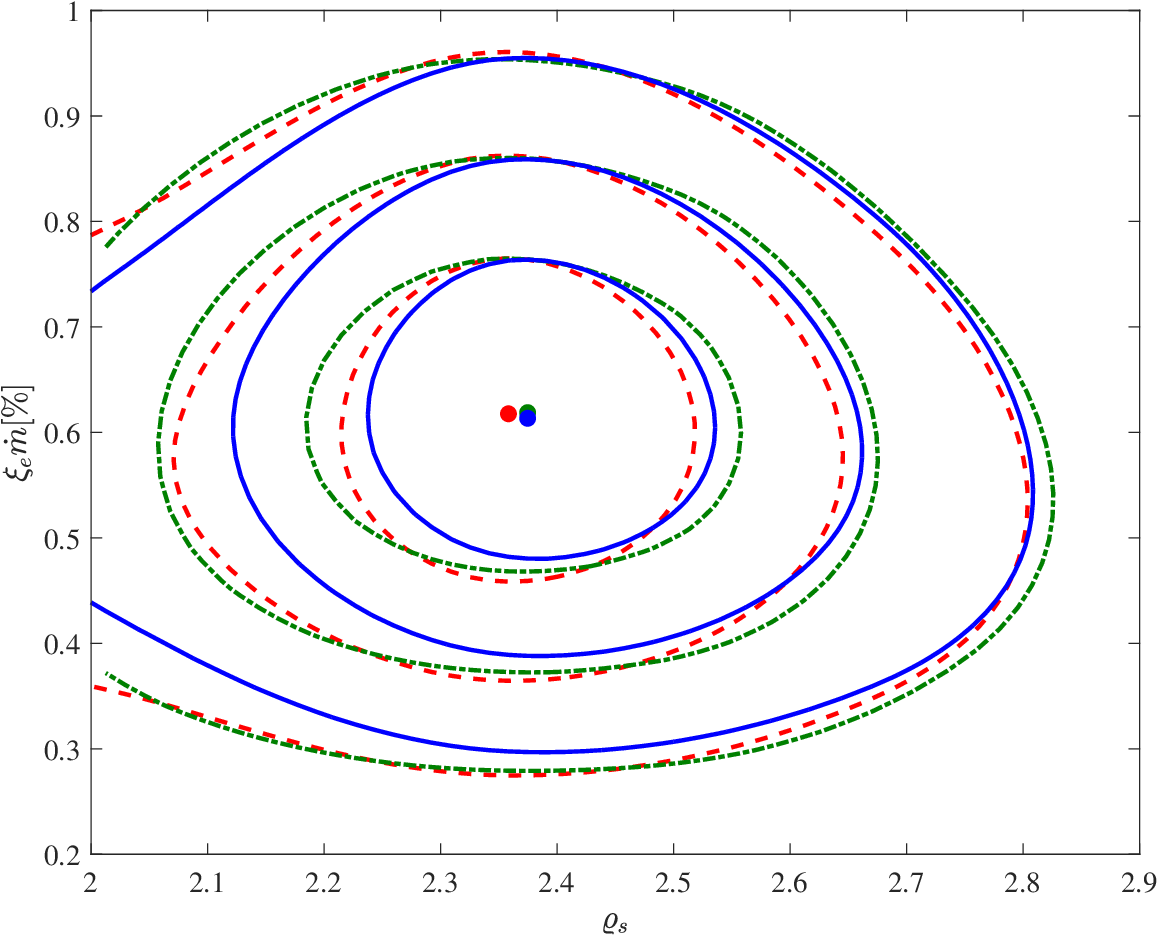}}
	\caption{\label{fig:healpix10CL}
		Two-parameter, $1\sigma\till3\sigma$ confidence intervals of the ring model parameters, CRE injection rate $\xi_e \dot{m}$ and scaled shock radius $\mytaur_v$, shown for $N_{hp}=9$ (solid blue contours), $N_{hp}=10$ (dot-dashed green contours), and $N_{hp}=11$ (dashed red contours).
	}
\end{figure}

\RefMarkA{
In Figure \ref{fig:healpix10}, the nominal results with HEALPix order $N_{hp}=10$ are compared with orders $N_{hp}=9$ and $11$, \ie using 4 times less or more than the nominal number of pixels.
The confidence level contours of the ring model parameters are shown to be converged for these $N_{hp}$ values in Figure \ref{fig:healpix10CL}.}

\RefDelB{
In Figure \ref{fig:check_normal}, we examine the distribution of the mock catalog results.
The $68\%$, $95\%$, $99.7\%$ confidence intervals from the mocks are compared to the $\pm1,2,3$ standard deviation intervals.
The agreement between these curves indicates that the mock distribution is consistent with a normal distribution. Using the mock catalogs' standard deviation would change the nominal significance by $<2\%$.}

\RefMarkB{
\subsection{Cluster sample selection tests}
}

\RefMarkB{
Our results are found to be robust to small changes in the cuts used to create the sample, such as changing the latitude cuts in the range $15\dgr\till25\dgr$, the $\theta_{500}$ minimal cut in the range $0.15\dgr-0.25\dgr$, the point source containment angle avoidance in the range $86\%-93\%$, and changing the $\theta_{500}$ maximal cut to $0.45\dgr$ (thus removing another two clusters).
}

\begin{figure}[t]
	\centerline{\epsfxsize=9.4cm \epsfbox{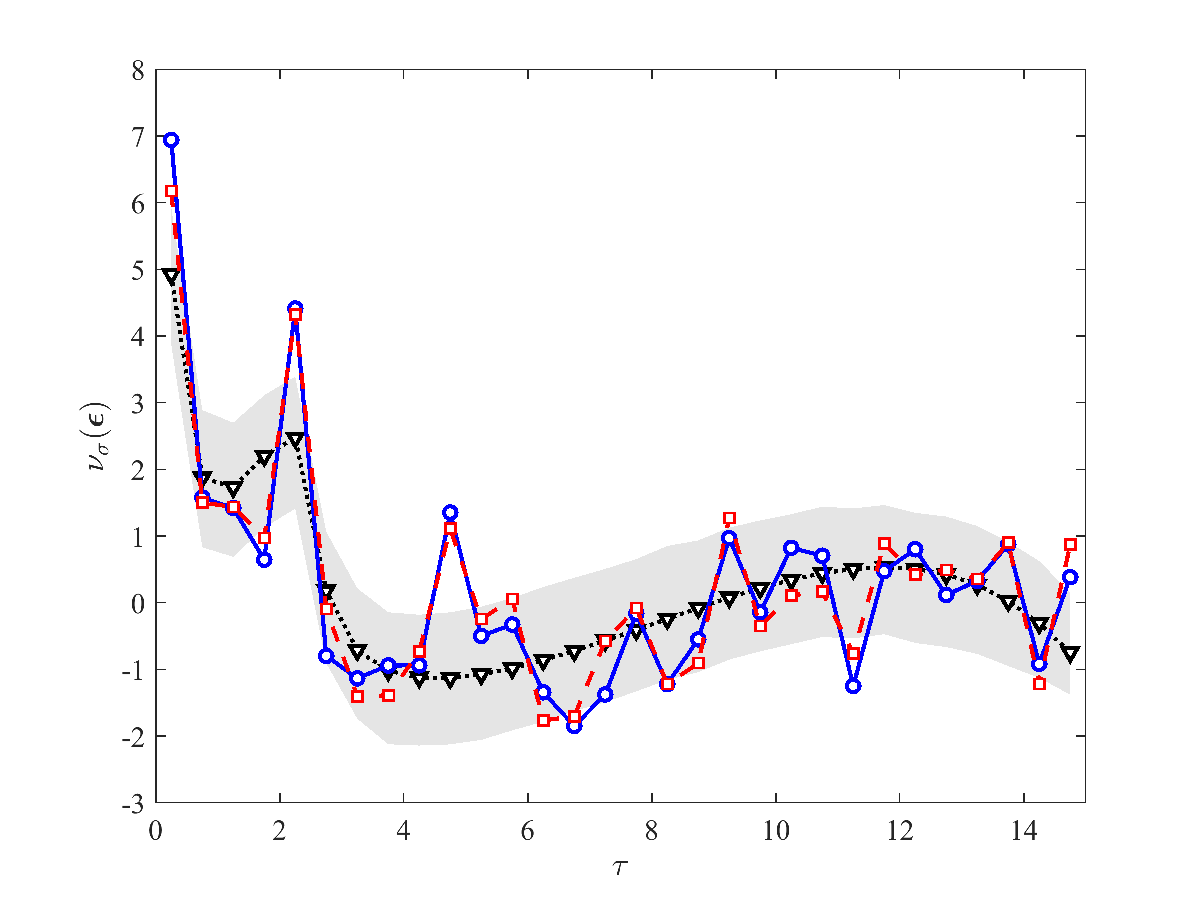}}
	\caption{\label{fig:dip_removed} The significance $\nu_\sigma(\bm{\epsilon})$ of the energy co-added excess \gama-ray counts over the foreground, shown as a function of $\mytautheta$ for the photon co-addition method (blue circles with solid lines to guide the eye) and for the cluster co-addition method (red squares with dashed lines to guide the eye), for the 96 clusters sub-sample without the two regions which are dense on the sky.
		Also shown is the simulated signals for the best fit model combining AGN with a spherical virial shock ($\xi_{e}\dot{m}=0.6\%, \myLAGN=1.6\times10^{41} \erg \se^{-1}$; black down-triangles with a dotted line and with the $1\sigma$ extent of the mock catalog distribution as a shaded region).
		}
\end{figure}

\RefMarkA{The angular separation between the clusters in our sample is on average sufficiently large to avoid an overlap between their individual regions of interest.
However, two regions on the sky show a high density of clusters, where the regions of interest overlap and some of the photons are double-counted.
One region contains 12 clusters around coordinates $\{l,b\}\simeq \{315\dgr,32\dgr\}$, and the other contains four clusters around coordinates $\{12\dgr,50\dgr\}$. Removing these 16 clusters from our sample strengthens (see Figure \ref{fig:dip_removed}) the $2<\tau<2.5$ signal to $\sim4.4\sigma$, while diminishing the $\tau<0.5$ signal by $\sim0.2\sigma$\RefMarkB{; neither effect is strong}. The change in the ring significance arises, in part, from the removal of analysis artifacts: overlapping virial rings contaminate the foreground of their host clusters, leading to errors in the modeled foreground.
This change may also have a physical origin: clusters at close proximity to each other could have more elongated virial shocks, diminishing the spherical component picked up by our radial binning.}

\RefMarkB{
Another outcome of removing the overlapping clusters is the disappearance of the $\sim(-2.5)\sigma$ dip seen in Figure \ref{fig:combined1} just outside the peripheral signal (in the $3<\tau<3.5$ bin), leaving an insignificant ($\sim0\sigma$) deviation from the foreground (see Figure \ref{fig:dip_removed}). In any case, this dip is not highly significant, even if all clusters are retained, especially when correcting for trial factors (see $\chi^2$ values in Table \ref{tab:fit_res}).
Nevertheless, we examine the origin of this putative feature. Although the origin of the dip is not fully understood, it can be traced in part to the amplification of double-counted fluctuations. Foreground removal in an overlap region is unlikely to generate the effect, as the latter remains intact for low-order, and even a constant, foreground removal (see Figure \ref{fig:diff Fg}). Moreover, a simple simulation based on the actual positions of the clusters on the sky does not reproduce the dip.
As the feature is insignificant and is entirely associated with overlap regions, we dismiss it.
}

\RefMarkB{
\subsection{Point source masking tests}
}

\RefMarkB{
The significance values of the central, point-like, excess emission and the
peripheral, ring-like, excess emission, change little when varying the masking containment area in the $83\%-93\%$ range, equivalent to a change of up to $\sim30\%$ in the masking angular radius. }

\RefMarkA{
The sensitivity of the results to sub-threshold, undetected point sources is tested by using the 2-year LAT point source catalog (2FGL) \cite{Fermi2FGL} for masking, instead of 3FGL. This earlier catalog was created using data with approximately half the exposure of 3FGL. Using 2FGL yields only slight changes in the significance of both the peripheral signal (no effect for the cluster co-addition, and a $\sim0.2\sigma$ reduction for the photon co-addition) and the central signal (a $\sim0.1\sigma$ reduction).
These results confirm that our conclusions are not sensitive to unresolved point sources.
}

\RefMarkB{
\subsection{Angular rescaling tests}
}

\RefMarkA{
By examining cluster sub-samples with increasingly narrower ranges of $\theta_{500}$, we test for systematic effects that may arise from the radial $\mytautheta$ rescaling, and for the statistical behavior of the stacked signals. We find that the significance \RefMarkB{values} of the signals indeed scale as expected from the statistics of the number of clusters in the sub-sample. For example, for the peripheral signal, where the full 112 cluster sample gives $\nu_\sigma(\bm{\epsilon})\simeq4.2$, the 83 clusters with $0.2\dgr<\theta_{500}<0.3\dgr$ give $\nu_\sigma(\bm{\epsilon})\simeq3.7$, and the 62 clusters with $0.2\dgr<\theta_{500}<0.25\dgr$ give $\nu_\sigma(\bm{\epsilon})\simeq3.0$, as expected from Poisson statistics.
This confirms that the rescaling of the cluster angular sizes does not introduce significant spurious systematic effects, and that the stacked signals arise from the cumulative contribution of many clusters, and are statistically limited.}

\RefDelB{
In Figure \ref{fig:dip_removed}, we show the significance of excess emission for the 96 clusters in the sub-sample in which the two densest cluster regions on the sky were removed to avoid photon double-counting. As the figure indicates, the central signal is slightly diminished in this sub-sample, while the peripheral signals is slightly enhanced. Another outcome of avoiding overlapping regions of interest is the removal of the dip-like feature in the $3<\tau<3.5$ bin, seen in Figures \ref{fig:combined1} and \ref{fig:combined2}.}

\RefMarkB{
\subsection{The central source effect on the ring signal}
}

We test for possible effects of the central sources on the significance of the peripheral, ring signal, in two methods.
In the first method, we mask an increasingly larger inner region (only) when computing the TS.
When masking the central, $\mytautheta<1.5$ parts of the clusters instead of modeling the central source, the TS-based significance declines with respect to the nominal $5.8\sigma$, as expected from the omission of data points, but remains $>4.5\sigma$ \RefMarkB{($\myTS>25$, see Table \ref{tab:fit_res})}; masking a smaller region ($\mytautheta<1$) gives an intermediate, $>4.8\sigma$ significance.
The (ring) parameters evaluated in this method change by $<10\%$ with respect to their nominal values. \RefMarkB{Even using only the $2<\tau<2.5$ bin gives a high, $>4\sigma$ $(\myTS>19)$, results.}
These results indicate that the central sources do not dominate the peripheral signal.
As a second, independent test, we repeat the analysis after removing clusters that show a significant central source, selected as $\nu_{\sigma,c}(\bm{\epsilon},\mytautheta<0.5)>2\sigma$.
This leaves a clean sub-sample of $93$ clusters, with no stacked central excess, $\nu_\sigma(\bm{\epsilon},\mytautheta<0.5)\simeq0$, yet showing a \RefMarkB{moderately} high, \RefDelB{$5.7\sigma$}\RefMarkB{$3.4\sigma$ $(\myTS=14.9)$} TS-based significance for the ring. In this test, $\mytaur_s$ changes by $<10\%$, but $\xi_e$ declines by $\sim 35\%$, partly because bright rings that contribute to the central bin were preferentially omitted.

\section{Cluster sample}
\label{sec:ClusterSample}

The clusters in our sample and their parameters are listed in Table \ref{tab:clusters} below.

\clearpage

{\scriptsize
	\begin{longtable}{llcccccccccc}
		\caption{\label{tab:clusters}
			MCXC Clusters used in this study, sorted by mass, with $\beta$-model parameters, where available. \textbf{Columns:} (1) Cluster catalog Name; (2) Alternate cluster name; (3) Galactic latitude; (4) Galactic longitude; (5) Mass within $\myR$ in $10^{14}M_\odot$ units; (6) Radius enclosing an overdensity $\delta=500$ in \kpc; (7) Angular scale of $\myR$ in degrees; (8) Cluster temperature in \keV; (9) The $\beta$-model $\beta$ parameter; (10) The $\beta$-model central electron density in units of $10^{-3}\cm^{-3}$; (11) The $\beta$-model core radius in \kpc; (12) Reference number for $\beta$-model parameters.
			}\\
		\hline
		MCXC Name &  Alt. Name &  b&  l &  $M_{500}$ &    $\myR$  &  $\theta_{500}$ &  $T$  &  $\beta$ &  $n_e$ &  $r_c $ &  Ref. \\
		(1) & (2) & (3) & (4) & (5) & (6) & (7) & (8) & (9) & (10) & (11)
        & (12) \\
		\hline
		\endfirsthead
		\hline
		MCXC Name &  Alt. Name &  b&  l &  $M_{500}$ &    $\myR$  &  $\theta_{500}$ &  $T$  &  $\beta$ &$n_e$&  $r_c $ &  Ref. \\
		\hline
		\endhead
		MCXC J0458.9-0029 & NGC1713 & -24.56 & 198.77 & 0.11 & 341 & 0.31 &    &   &    &   & \\	
		MCXC J0113.9-3145 & S141A & -83.27 & 257.67 & 0.11 & 341 & 0.24 &    &   &    &   & \\	
		MCXC J1336.6-3357 & A3565 & 27.98 & 313.54 & 0.13 & 354 & 0.39 & 0.87 & 0.31  & 1.39 & 32  & \Fukazawa\\
		MCXC J0920.0+0102 & MKW 1S & 33.06 & 230.94 & 0.13 & 359 & 0.28 &    &   &    &   & \\
		MCXC J1329.4+1143 & MKW 11 & 72.20 & 334.86 & 0.17 & 388 & 0.23 &    &   &    &   & \\
		MCXC J1257.1-1339 & RXC J1257.1-1339  & 49.19 & 305.06 & 0.18 & 399 & 0.36 &    &   &    &   & \\
		MCXC J1751.7+2304 & NGC 6482 & 22.92 & 48.09 & 0.19 & 403 & 0.41 &    &   &    &   & \\
		MCXC J2035.7-2513 & A3698 & -33.24 & 19.25 & 0.20 & 413 & 0.28 &    &   &    &   & \\
		MCXC J0933.4+3403 & UGC 05088 & 47.24 & 191.05 & 0.21 & 413 & 0.21 &    &   &    &   & \\
		MCXC J1304.2-3030 & RXC J1304.2-3030 & 32.27 & 306.20 & 0.23 & 428 & 0.50 &    &   &    &   & \\
		MCXC J1723.3+5658 & NGC 6370 & 34.34 & 85.21 & 0.23 & 428 & 0.22 &    &   &    &   & \\
		MCXC J1736.3+6803 & ZW 1745.6+6703 & 32.00 & 98.27 & 0.24 & 435 & 0.24 & 1.37 & 0.38 & 4.39 & 13 & \Fukazawa\\
		MCXC J2347.4-0218 & HCG 97 & -60.83 & 88.50 & 0.24 & 438 & 0.27 &  1.2 & 0.4 & 8.46 & 7 & \Fukazawa \\
		MCXC J1334.3+3441 & NGC 5223 & 78.09 & 74.98 & 0.25 & 442 & 0.25 &    &   &    &   & \\
		MCXC J1847.3-6320 & S0805 & -23.60 & 332.25 & 0.26 & 448 & 0.42 &    &   &  &  & \\
		MCXC J1253.0-0912 & HCG 62 & 53.67 & 303.62 & 0.27 & 455 & 0.42 &  1.1 & 0.4 & 23.18 & 5 & \Fukazawa\\
		MCXC J2249.2-3727 & S1065 & -62.35 &  3.29 & 0.31 & 472 & 0.23 &    &   &    &   & \\
		MCXC J1755.8+6236 & RXC J1755.8+6236 & 30.22 & 91.82 & 0.31 & 474 & 0.25 & 1.78 & 0.38 & 1.99 & 35 & \Fukazawa \\
		MCXC J2214.8+1350 & RX J2214.7+1350 & -34.13 & 75.17 & 0.32 & 477 & 0.26 &    &   &    &   & \\
		MCXC J2250.0+1137 & RX J2250.0+1137 & -41.34 & 81.71 & 0.32 & 479 & 0.26 &    &   &    &   & \\
		MCXC J0838.1+2506 & CGCG120-014 & 33.73 & 199.58 & 0.33 & 483 & 0.23 &    &   &    &   & \\
		MCXC J0916.1+1736 &   & 39.68 & 211.99 & 0.33 & 483 & 0.23 &    &   &    &   & \\
		MCXC J1606.8+1746 & A2151B & 44.17 & 31.82 & 0.33 & 483 & 0.21 &    &   &    &   & \\
		MCXC J1615.5+1927 & NGC 6098 & 42.81 & 34.97 & 0.36 & 499 & 0.22 &    &   &    &   & \\
		MCXC J2111.6-2308 & AM2108 & -40.49 & 24.69 & 0.37 & 500 & 0.21 &    &   &    &   & \\
		MCXC J1329.5+1147 & MKW 11 & 72.24 & 335.02 & 0.38 & 508 & 0.32 &    &   &    &   & \\
		MCXC J0907.8+4936 & VV 196 & 42.12 & 169.27 & 0.40 & 513 & 0.20 &    &   &    &   & \\
		MCXC J0125.6-0124 & A0194 & -63.00 & 142.07 & 0.40 & 516 & 0.39 &  1.9 & 0.4 & 1.11 & 81 & \Fukazawa\\
		MCXC J1050.4-1250 & USGC S152 & 40.40 & 262.76 & 0.41 & 522 & 0.46 &    &   &   &  & \\
		MCXC J1206.6+2811 & MKW4S & 80.02 & 204.27 & 0.42 & 523 & 0.26 & 1.90 & 0.38 & 5.32 & 18  & \Fukazawa \\
		MCXC J0712.0-6030 & AM 0711 & -20.94 & 271.26 & 0.43 & 529 & 0.23 &    &   &    &   & \\
		MCXC J0249.6-3111 & S0301 & -63.96 & 229.00 & 0.45 & 536 & 0.32 &    &   &    &   & \\
		MCXC J2101.5-1317 &   & -34.80 & 35.31 & 0.45 & 535 & 0.26 &    &   &    &   & \\
		MCXC J1109.7+2145 & A1177 & 66.28 & 220.43 & 0.46 & 540 & 0.24 &    &   &    &   & \\
		MCXC J0036.5+2544 & ZWCL193 & -37.01 & 118.75 & 0.49 & 549 & 0.22 &    &   &    &   & \\
		MCXC J2224.7-5632 & S1020 & -50.71 & 334.28 & 0.50 & 552 & 0.22 &    &   &    &   & \\
		MCXC J1840.6-7709 & RXC J1840.6-7709  & -25.80 & 317.21 & 0.54 & 571 & 0.40 &    &   &    &   & \\
		MCXC J2315.7-0222 & NGC 7556 & -56.28 & 76.06 & 0.58 & 585 & 0.30 &    &   &    &   & \\
		MCXC J0624.6-3720 & A3390 & -21.05 & 245.14 & 0.60 & 589 & 0.25 &    &   &    &   & \\
		MCXC J0454.8-1806 & CID 28 & -33.63 & 217.45 & 0.62 & 595 & 0.25 &  1.8 & 0.5 & 5.64 & 35 & \Fukazawa \\
		MCXC J0150.7+3305 & A260 & -28.16 & 137.01 & 0.63 & 597 & 0.23 &    &   &    &   & \\
		MCXC J0228.1+2811 & RX J0228.2+2811 & -30.01 & 147.57 & 0.66 & 608 & 0.24 &    &   &    &   & \\
		MCXC J2107.2-2526 & A3744 & -40.14 & 21.44 & 0.68 & 612 & 0.23 &    &   &    &   & \\
		MCXC J2043.2-2629 & S0894 & -35.22 & 18.35 & 0.69 & 616 & 0.21 &    &   &    &   & \\
		MCXC J0110.0-4555 & A2877 & -70.85 & 293.05 & 0.71 & 625 & 0.36 &    &   &    &   & \\
		MCXC J1440.6+0328 & MKW 8 & 54.79 & 355.49 & 0.74 & 632 & 0.33 &    &   &  &   & \\
		MCXC J0113.0+1531 & A0160 & -47.03 & 130.60 & 0.74 & 631 & 0.20 &    &   &    &   & \\
		MCXC J0058.9+2657 & RX J0058.9+2657 & -35.89 & 124.99 & 0.82 & 651 & 0.20 &    &   &    &   & \\
		MCXC J1331.5-3148 & RX J1331.5-3148 & 30.29 & 312.80 & 0.83 & 655 & 0.21 &    &   &    &   & \\
		MCXC J0542.1-2607 & CID 36 & -26.02 & 230.42 & 0.85 & 661 & 0.24 &    &   &    &   & \\
		MCXC J1733.0+4345 & IC 1262 & 32.07 & 69.52 & 0.86 & 664 & 0.30 &    &   &    &   & \\
		MCXC J1740.5+3538 & RX J1740.5+3539 & 29.07 & 60.60 & 0.91 & 676 & 0.22 &    &   &    &   & \\
		MCXC J2101.8-2802 & A3733 & -39.60 & 17.77 & 0.92 & 679 & 0.25 &    &   &    &   & \\
		MCXC J1253.2-1522 & A1631 & 47.49 & 303.57 & 0.98 & 692 & 0.21 &  2.28 & 0.85  & 0.49  & 15 & \Fukazawa \\
		MCXC J0040.0+0649 & A76 & -55.94 & 117.86 & 0.99 & 695 & 0.25 &    &   &    &   &  \\
		MCXC J1407.4-2700 & A3581 & 32.86 & 323.14 & 1.08 & 719 & 0.43 &  1.7 & 0.5 & 40.60 & 9& \Fukazawa \\
		MCXC J0525.5-3135 & A3341 & -31.09 & 235.17 & 1.09 & 718 & 0.26 &      &     &      &    &  \\
		MCXC J0341.2+1524 & IIIZw54 & -30.79 & 172.18 & 1.13 & 728 & 0.33 & 2.16 & 0.887 & 2.42 & 198 & \Chen \\
		MCXC J0828.6+3025 & A0671 & 33.15 & 192.75 & 1.15 & 728 & 0.21 &      &     &      &    &  \\
		MCXC J2310.4+0734 & Pegasus II & -47.54 & 84.15 & 1.17 & 735 & 0.24 &      &     &      &    &  \\
		MCXC J0003.2-3555 & A2717 & -76.49 & 349.33 & 1.20 & 739 & 0.21 &      &     &      &    &  \\
		MCXC J2338.4+2700 & A2634 & -33.09 & 103.48 & 1.22 & 746 & 0.34 &  3.5 & 0.4 & 2.17 & 75& \Fukazawa \\
		MCXC J0115.2+0019 & A0168 & -61.95 & 135.65 & 1.25 & 749 & 0.24 &      &     &      &    &  \\
		MCXC J0500.7-3840 & A3301 & -37.41 & 242.41 & 1.27 & 752 & 0.20 &      &     &      &    &  \\
		MCXC J0108.8-1524 & A0151 & -77.60 & 142.84 & 1.28 & 753 & 0.20 &      &     &      &    &  \\
		MCXC J0025.5-3302 & S0041 & -81.85 & 344.77 & 1.29 & 756 & 0.22 &      &     &      &    &  \\
		MCXC J1604.5+1743 & A2151 & 44.66 & 31.48 & 1.32 & 765 & 0.29 &  2.1 & 0.5 & 9.98 & 35& \Fukazawa \\
		MCXC J2104.9-5149 & RXC J2104.9-5149 & -41.38 & 346.39 & 1.32 & 763 & 0.22 &      &     &      &    &  \\
		MCXC J1811.0+4954 & ZwCl 8338 & 26.71 & 77.72 & 1.35 & 767 & 0.22 &      &     &      &    &  \\
		MCXC J0548.6-2527 & A0548E & -24.42 & 230.26 & 1.38 & 776 & 0.26 &      &     &      &    &  \\
		MCXC J2324.3+1439 & A2593 & -43.18 & 93.45 & 1.42 & 783 & 0.26 &      &     &      &    &  \\
		MCXC J1539.6+2147 & A2107 & 51.53 & 34.40 & 1.49 & 796 & 0.27 &  3.5 & 0.6 & 5.76 & 85&  \\
		MCXC J1255.5-3019 & A3530 & 32.53 & 303.99 & 1.56 & 804 & 0.21 &  3.7 & 0.4 & 4.93 & 37&  \\
		MCXC J2018.7-5242 & S0861 & -34.28 & 345.83 & 1.58 & 809 & 0.23 &      &     &      &    &  \\
		MCXC J0246.0+3653 & A0376 & -20.55 & 147.11 & 1.61 & 815 & 0.24 &  3.7 & 0.6 & 3.09 & 114&  \\
		MCXC J2113.8+0233 & IC 1365 & -29.83 & 53.51 & 1.64 & 820 & 0.24 &  3.4 & 0.7 & 1.69 & 250&  \\
		MCXC J1454.5+1838 & A1991 & 60.50 & 22.79 & 1.68 & 823 & 0.20 &      &     &      &    &  \\
		MCXC J0721.3+5547 & A0576 & 26.25 & 161.36 & 1.68 & 829 & 0.31 &  3.7 & 0.6 & 4.02 & 101& \Fukazawa \\
		MCXC J1332.3-3308 & A3560 & 28.95 & 312.72 & 1.68 & 827 & 0.24 & 3.16 & 0.566 & 2.05 & 175 & \Chen \\
		MCXC J1329.7-3136 & SC1327-312 & 30.56 & 312.40 & 1.72 & 833 & 0.24 & 3.42 & 0.34 & 2.71 & 64 & \Fukazawa \\
		MCXC J0125.0+0841 & A193 & -53.27 & 136.92 & 1.76 & 839 & 0.24 &      &     &      &    &  \\
		MCXC J2227.8-3034 & A3880 & -58.51 & 18.00 & 1.79 & 841 & 0.21 &      &     &      &    &  \\
		MCXC J2235.6+0128 & A2457 & -46.59 & 68.63 & 1.82 & 846 & 0.20 &      &     &      &    &  \\
		MCXC J1017.3-1040 & A0970 & 36.86 & 253.05 & 1.85 & 850 & 0.21 &      &     &      &    &  \\
		MCXC J2344.9+0911 & A2657 & -50.26 & 96.72 & 1.88 & 859 & 0.30 &  3.9 & 0.5 & 6.34 & 66& \Fukazawa \\
		MCXC J1252.5-3116 & RBS 1175  & 31.60 & 303.22 & 1.89 & 858 & 0.23 &      &     &      &    &  \\
		MCXC J1359.2+2758 & A1831 & 74.95 & 40.07 & 1.98 & 869 & 0.20 &      &     &      &    & \\
		MCXC J2347.7-2808 & A4038 & -75.86 & 25.14 & 2.04 & 886 & 0.41 &  2.9 & 0.5 & 34.53 & 14& \Fukazawa \\
		MCXC J0011.3-2851 & A2734 & -80.99 & 19.56 & 2.06 & 881 & 0.20 & 3.85 & 0.624 & 3.87 & 145 & \Chen \\
		MCXC J0330.0-5235 & A3128 & -51.12 & 264.80 & 2.08 & 883 & 0.20 &      &     &      &    &  \\
		MCXC J0425.8-0833 & RBS 0540 & -36.16 & 203.30 & 2.09 & 891 & 0.31 &      &     &      &    &  \\
		MCXC J1523.0+0836 & A2063 & 49.68 & 12.81 & 2.16 & 902 & 0.35 &  3.4 & 0.6 & 5.78 & 95& \Fukazawa \\
		MCXC J0351.1-8212 & S0405 & -32.49 & 296.42 & 2.19 & 899 & 0.21 &  4.2 & 0.7 & 1.45 & 314& \Fukazawa \\
		MCXC J1257.2-3022 & A3532 & 32.48 & 304.43 & 2.34 & 920 & 0.24 &  4.3 & 0.5 & 3.84 & 102& \Fukazawa \\
		MCXC J1333.6-3139 & A3562 & 30.36 & 313.33 & 2.37 & 927 & 0.27 &  4.5 & 0.4 & 12.40 & 29& \Fukazawa \\
		MCXC J1516.7+0701 & A2052 & 50.12 &  9.41 & 2.49 & 947 & 0.37 &  3.0 & 0.5 & 34.63 & 21& \Fukazawa \\
		MCXC J1521.8+0742 & MKW 3S & 49.46 & 11.39 & 2.52 & 947 & 0.30 &  3.2 & 0.6 & 21.98 & 36& \Fukazawa \\
		MCXC J2357.0-3445 & A4059 & -76.08 & 356.36 & 2.67 & 964 & 0.29 &  4.0 & 0.6 & 8.30 & 85& \Fukazawa \\
		MCXC J1326.9-2710 & A1736 & 35.02 & 312.57 & 2.71 & 969 & 0.30 &  3.2 & 0.5 & 1.97 & 125& \Fukazawa \\
		MCXC J0338.6+0958 & 2A0335 & -35.05 & 176.26 & 3.45 & 1055 & 0.42 &  3.1 & 0.5 & 117.8 & 11& \Fukazawa \\
		MCXC J0918.1-1205 & A0780 & 25.10 & 242.93 & 3.62 & 1066 & 0.28 &  3.6 & 0.6 & 15.62 & 66& \Fukazawa \\
		MCXC J0342.8-5338 & A3158 & -48.93 & 265.05 & 3.65 & 1067 & 0.26 &  5.0 & 0.6 & 6.17 & 108& \Fukazawa \\
		MCXC J1327.9-3130 & A3558 & 30.73 & 311.99 & 3.97 & 1101 & 0.33 &  5.3 & 0.5 & 5.68 & 113& \Fukazawa \\
		MCXC J0257.8+1302 & A0399 & -39.46 & 164.32 & 4.25 & 1117 & 0.23 &  6.4 & 0.6 & 4.53 & 142& \Fukazawa \\
		MCXC J1703.8+7838 & A2256 & 31.76 & 111.01 & 4.25 & 1122 & 0.28 &  6.5 & 0.9 & 2.67 & 400& \Fukazawa \\
		MCXC J0317.9-4414 & A3112 & -56.08 & 252.93 & 4.39 & 1129 & 0.22 &  4.5 & 0.6 & 37.01 & 32& \Fukazawa \\
		MCXC J2012.5-5649 & A3667 & -33.39 & 340.86 & 5.17 & 1199 & 0.31 &  5.9 & 0.5 & 4.22 & 130& \Fukazawa \\
		MCXC J1348.8+2635 & A1795 & 77.18 & 33.82 & 5.53 & 1224 & 0.28 &  5.8 & 0.6 & 15.62 & 83& \Fukazawa \\
		MCXC J0258.9+1334 & A0401 & -38.87 & 164.18 & 5.85 & 1242 & 0.25 &  8.0 & 0.7 & 5.64 & 207& \Fukazawa \\
		MCXC J0413.4+1028 & A0478 & -28.29 & 182.43 & 6.42 & 1276 & 0.21 &  7.6 & 0.6 & 49.00 & 48& \Fukazawa \\
		MCXC J1510.9+0543 & A2029 & 50.53 &  6.44 & 7.27 & 1334 & 0.26 &  7.5 & 0.6 & 24.17 & 75& \Fukazawa \\
		MCXC J1558.3+2713 & A2142 & 48.69 & 44.22 & 8.15 & 1380 & 0.23 &  9.3 & 0.7 & 12.30 & 139& \Fukazawa \\

	\end{longtable}
}

\acknowledgments
We thank J. Mushkin, I. Gurwich, Y. Lyubarsky, A. Zitrin, E. Ofek,  D. Prokhorov, E. Charles and O. Reimer for helpful discussions.
This research has received funding from the IAEC-UPBC joint research foundation (grants No. 257/14 and 300/18), and was supported by
the Israel Science Foundation (grant No. 1769/15) and by the GIF (grant I-1362-303.7/2016).


\providecommand{\href}[2]{#2}\begingroup\raggedright\endgroup

\end{document}